\documentclass[11pt]{article}
\usepackage{hyperref}
\usepackage{authblk}
\usepackage{graphicx}
\usepackage{caption}
\usepackage{amsmath}
\usepackage{amssymb}
\usepackage{mathtools}
\usepackage{multirow}
\usepackage{shrthnds}
\usepackage{amsmath}
\usepackage{color}
\usepackage{subcaption}
\usepackage{hyperref}
\usepackage{wasysym}
\usepackage{ulem}

\title{ \bf Cosmological Dark Matter in a Conformal Model }
\author{Prasenjit Sanyal$^{1,}$\footnote{ psanyal@iitk.ac.in}}
\author{Alekha C. Nayak$^{2,}$\footnote{ alekhanayak@nitm.ac.in}}
\author{Gopal Kashyap$^{3,}$\footnote{ gplkumar87@gmail.com }}
\author{Pankaj Jain$^{1,}$\footnote{ pkjain@iitk.ac.in}}
\affil{
$^1$Department of Physics, Indian institute of Technology, Kanpur 208016, India\\
$^2$ Department of Physics, National Institute of Technology, Meghalaya, Shillong, Meghalaya 793003, India\\
$^3$Department of Physics, School of Basic and Applied Sciences, Galgotias University, Greater Noida, India}
\begin{document}
\maketitle

\begin{abstract}
We study the collider, astrophysical and cosmological 
constraints on the dark matter sector of
a conformal model within the framework of both the freeze out as well as the
freeze in mechanism. 
The model has a dark sector with strong self 
interactions. 
This sector couples weakly 
 with the Standard Model (SM) particles via a scalar messenger. The lightest 
dark sector particle is a pion-like fermion anti-fermion bound state. 
We find that the model successfully satisfies the constraints coming from the Higgs decay 
to the visible as well as the invisible sector. 
 We have used the results
of the dark matter direct detection experiments, such as,
 XENON1T in order to impose bounds on the parameters of the model. 
The model satisfies the indirect detection constraints of gamma ray from the galactic center and Dwarf spheroidal galaxies.
We also determine the parameter range for which it satisfies the 
 astrophysical constraints on the dark matter self coupling.     
\end{abstract}

\clearpage
\section{Introduction}
\label{intro}

  The presence of non-baryonic dark matter (DM) in the Universe is very well established by cosmological and astrophysical observations. 
The Planck and WMAP data suggests that dark matter density  $\Omega_Xh^2=0.1187\pm 0.0017$ \cite{Ade:2015xua}. Currently there are many ground and satellite based experiments looking 
for direct evidence of dark matter. The direct detection experiments
of dark matter are based on elastic scattering of DM with the detector 
nucleons. No evidence of DM detection is obseved till date and there exist rather stringent upper limits on
the cross section of dark matter with nucleons from LUX 
 \cite{Akerib:2016vxi,Akerib:2015rjg}, PandaX-II \cite{Cui:2017nnn,Tan:2016zwf}, XENON1T \cite{Aprile:2018dbl}, SuperCDMS \cite{Agnese:2015nto} and CRESST-II \cite{Angloher:2015ewa}.
Indirect detection of DM comes through its 
decay or annihilation at the center of galaxies.
Despite several tentative claims of detection, such as, the 1-3 GeV gamma ray excess emission from the Galactic center
by Fermi-LAT collaboration \cite{Daylan:2014rsa}, so far there does not exist conclusive evidence for 
a dark matter candidate 
 \cite{TheFermi-LAT:2017vmf}.  
Apart from that DM may annihilate or decay into monoenergetic $\gamma-$rays. Search for monoenergetic spectral lines in the $Fermi$ Large Area Telescope ($Fermi$-LAT) observations produced null results for the DM 
in the range of 200 MeV - 500 GeV. The non-observation of such a spectral line puts an upper bound (95$\%$CL) on the annihilation cross section
 and decay widths of DM candidates \cite{Ackermann:2015lka}. Joint analysis of MAGIC Cherenkov telescopes and $Fermi$-LAT \cite{Ahnen:2016qkx} on gamma ray data from Dwarf spheroidal galaxies (dSphs) also imposes an upper bound on DM annihilation cross-section in the mass range of 10 GeV to 100 TeV.

%There also exists an interesting observation of 3.5 keV X-ray line which has been explained earlier in terms of dark matter annihilation or decay \cite{Bulbul:2014sua,Boyarsky:2014jta,Babu:2014pxa}. However this now  appears to be of astrophysical origin and we do not consider it further in our paper. 

In this paper we consider a conformal model discussed earlier 
\cite{Meissner:2006zh,Meissner:2007xv,Hur:2011sv,Jain:2015aoa,Jain:2014wsa}. 
This model has self interacting dark matter sector which is assumed
to have a QCD like dynamics. The action has conformal symmetry 
and the dark sector communicates with the electroweak sector through
a real scalar field $\chi$. The dynamical 
symmetry breaking of the dark matter sector leads to a vacuum expectation
value (VEV) of $\chi$ which in turn induces the   
electroweak symmetry breaking. In order to handle the strong interaction
dynamics, we consider an  
 effective action expressed in terms of dark sector pions $\Pi^i$, nucleons and
 scalar fields $\Sigma$ and $\chi$. The dark pion acts as the dark matter candidate. The scalar field $\chi$ acts as the messenger field through which the dark
pions interact with the Standard Model (SM) sector and $\Sigma$ is the bound state scalar. The three scalar fields, $\Phi$, $\chi$ and $\Sigma$ 
mix with one another. Here $\Phi$ is the Standard Model scalar field which
would be equivalent to the Higgs in the 
absence of mixing with dark sector scalars. After 
 diagonalizing their mass matrix we obtain three physical scalars, $\chi_1, \chi_2$ and $\chi_3$. We identify $\chi_1$ as the SM Higgs 
 and $\chi_2$ is a massive particle with its mass proportional to the symmetry breaking scale of dark sector. 
The $\chi_3$ particle is classically massless but acquires a mass
in quantum theory due to conformal anomaly. We discuss this in more detail
below. The 
dark sector particles $\chi_2$ and $\chi_3$ decay to the SM sector or to the 
dark pions within the lifespan of the Universe. 
For simplicity here we assume that there exists only a 
 single generation of dark fermions. In this case
we only have one dark pion which acts as a dark matter candidate.

% In our model the vev of $\chi$, $\eta$ is of the order $10^9$ or higher and we set the pion mass in the range 30-50GeV, also the Lagrangian parameters $\lambda_2$ and $\lambda_6$ being very small compared to 1, that the mass of $\chi_3$ becomes sufficiently larger than the mass of pions. Hence within the lifespan of universe $\chi_3$ decays to pions, apart from that $\chi_3$ can also decay to other SM sectors($b\overline{b}$ dominantly). Thus $\chi_3$ does not contribute into relic abundance. $\chi_2$ being massless, does not contribute in relic abundance.

 %In our model the proper relic density of pions is obtained for $\eta$ of the order $10^9$ or higher. 
 
We consider the cosmological implications of this model assuming both a 
freeze out and freeze in scenarios. As we shall see, within the 
freeze out scenario, there exists parameter range in which the model
satisfies all constraints including the constraint on 
$\sigma_{DM}/M_{\text{DM}}$ indicated by some astrophysical observations. Here
$\sigma_{DM}$ is the cross section for DM-DM scattering and $M_{\text{DM}}$ the mass
of dark matter particles. We next study the dark matter assuming the
freeze in scenario. We show that this model can accommodate to a very
low dark pion mass, as required for a dark matter candidate in this
case. In this case the astrophysical constraint on $\sigma_{DM}/M$ imposes 
some limits on the parameters of the model.

 The paper is organized as follows :  In section \ref{Review}, we review 
the conformal model and identify the different particle states predicted by this
model. 
In section \ref{sec:phenomenology} we determine the
 relic 
density of dark pions in the freeze-out scenario and also impose the collider, direct and indirect DM detection
constraints on the conformal model. 
In section \ref{sec:freezein} we determine the implications of dark matter
assuming the freeze in scenario.
In section \ref{sec:results} we show the final allowed space taking all the
constraints into account both for freeze out and freeze in scenarios. 
Finally we conclude in section \ref{sec:conclusions}.

\section{Review of the Conformal Model}
\label {Review}
The conformal model introduced in \cite{Meissner:2006zh,Meissner:2007xv,Hur:2011sv} has a strongly coupled dark matter 
sector similar to QCD. 
The action for the dark sector can be 
written as
\begin{equation}
  {\mc S}_D = \int d^4x \left[ -{1\over 4} G_{\mu\nu}^a G^{a\mu\nu }
  +i\bar\xi^i \gamma^\mu D_\mu\xi^i - g_Y^\chi\bar\xi^i\chi\xi^i \right],
  \label{eq:stronglycoupled}
\end{equation}
where $G_{\mu\nu}^a$ is the field strength tensor of the dark sector 
strong interaction mediator, $\xi^i$ represent
fermion fields and $\chi$ is a real scalar field. 
We refer to this strongly coupled sector as hypercolor
and call the quarks and gluons in this sector 
as hyperquarks and hypergluons. In our phenomenological study we  
consider only one multiplet of hypercolor fermions but in general there can 
be several multiplets. The hyperquarks and hypergluons 
will form bound state dark pions and nucleons. The dark pions may act as 
dark matter candidates. As we shall see they  
are able to satisfy all astrophysical and cosmological constraints. 
%\sout{In Eq. \ref{eq:stronglycoupled} we have also included a parity violating
%Yukawa interaction term proportional to the parameter $d$. This is required
%in order to have a non-zero decay rate of pions.}

The dark sector couples to the SM particles by the coupling
of $\chi$ to the Higgs sector. 
The action for the scalar sector of the model can be written as
\ba
\mathcal{S}_{S} &=& \int d^4x \Bigg[{1\over 2}g^{\mu\nu}\partial_\mu\chi
  \partial_\nu\chi+
   g^{\mu\nu} (D_\mu \mc H)^\dag(D_\nu \mc H) 
  - {\lambda_1\over 4}   \Big(2\mc H^\dag \mc H-\lambda_2\chi^2\Big)^2 
-{\lambda\over 4}\chi^4
\Bigg]
   \label{eq:S_EW_d}
   \ea
where $\mc H$ is the Higgs multiplet and $D_\mu$ is the SM gauge covariant
derivative. 
The Higgs multiplet can be decomposed as,
\begin{equation}
  \mc H = {1\over \sqrt{2}} \left(\begin{array}{c}\phi_1+i\phi_2\\
     \phi_3+i\phi_4 \end{array}\right). 
   \label{eq:Higgsmultiplet}
 \end{equation}
The action has conformal symmetry and we include all terms in the scalar
potential which are invariant under this symmetry.
In analogy with QCD the hyperquarks form condensates $\langle \overline{\xi}\xi\rangle $ with effective mass scale $\Lambda_S$.
 Once the condensate is generated we can substitute it in the dark sector Yukawa terms in the Lagrangian and minimize the potential  over the scalar fields $\phi_3$ and $\chi$. This generates non-zero values for the VEV of these fields and triggers electroweak symmetry breaking.
 We denote the VEV of  $\phi_3$ and $\chi$ as $v_{EW}$ and $\eta$ respectively.
Here $v_{EW}$ is the electroweak symmetry breaking scale.

 The model is interesting since it leads to 
a self interacting dark matter candidate which is preferred by cosmological
observations \cite{Spergel:1999mh,Bento:2001yk,Massey:2015dkw,Harvey:2015hha,Kahlhoefer:2015vua,Hochberg:2014kqa,Kaplinghat:2015aga}.  
The astrophysical implications of this conformal model have been studied
earlier in \cite{Heikinheimo:2014xza}. 
Due to scale symmetry we expect that 
classically the mass of one of the scalar particles is zero. However we 
expect this particle to acquire mass in the full quantum theory due 
to scale anomaly
\cite{Collins:1977}. We include this scale breaking by adding a term 
in the effective action  \cite{Gomm:1985ut}. 

The dark pions and nucleons are potential dark matter candidates. 
We use an effective model which is similar to the
 linear sigma model in order to handle
 these bound state fields.
The resulting dark sector effective Lagrangian can be
written as,
\begin{equation}
  {\cal L}_\Sigma =\bar \Psi i\gamma^\mu\partial_\mu\Psi
  +{1\over 2} \partial^\mu\Sigma\partial_\mu\Sigma
  +{1\over 2} \partial^\mu\Pi\partial_\mu\Pi
  - g_\Psi\bar \Psi (\Sigma+i \Pi\gamma_5)\Psi -
  {\lambda_5\over 4}\left(\Sigma^2+\Pi^2-\lambda_{6}\chi^2\right)^2\,.
  \label{eq:Sigmamodel}
\end{equation}
where $\Pi$,
$\Psi$ and $\Sigma$ 
 denote the dark sector pions, nucleons and the scalar fields
respectively. 
This model becomes same as the linear sigma model if we
replace the $\chi$ field with its vacuum expectation value. 
Here we have chosen this Lagrangian because it satisfies conformal 
invariance. 
So far the model has chiral symmetry which is not a symmetry of
the hypercolor interactions due to the presence of the Yukawa terms. We
break this symmetry by adding the 
following term,
\begin{equation}
  \mathcal{L}_\Pi = -{\lambda_7\over 2}\Pi^2\chi^2\,,
  \label{chi_break}
\end{equation}
With the addition of this term the dark sector pions acquire mass.

We next need to minimize the scalar potential given in Eqs.
\ref{eq:S_EW_d} and \ref{eq:Sigmamodel}. 
So far we have maintained
scale invariance in the effective action. However even though the original action
(Eq. \ref{eq:stronglycoupled}) is classically scale invariant,
the scale invariance is broken due to the anomaly \cite{Collins:1977}.
Hence the effective action which represents
the low energy dynamics of the bound states need not have
this symmetry. We also need to break this symmetry in order to obtain a
  non-trivial minimum of the potential without fine tuning any parameter
to zero \cite{Jain:2015aoa}.
We break the scale symmetry by modifying the potential
term for $\chi$ such that \cite{Gomm:1985ut}
\begin{equation}
 {\lambda\over 4} \chi^4 \rightarrow {\lambda\over 4} \chi^4\log\left({\chi^2\over 
\Lambda^2}\right) 
  \label{eq:scaleanomaly}
\end{equation}
where $\Lambda$ is a dimensional parameter related to the strong interaction
scale $\Lambda_S$.  
With this modification the potential acquires a non-trivial minimum
with non-zero vacuum expectation values for the fields $\Sigma$ and $\chi$. 
The scale
breaking terms lead to a mass term for the dilaton which
would be massless in the limit of exact scale invariance.

The minimum of the dark sector potential occurs at
\ba
\Sigma^2+\Pi^2 =\lm_6 \et^2
\ea
with
\ba
\la \chi \ra = \eta=\Lambda/\exp(1/4), \quad 
\la \phi_3 \ra = v_{EW}, \quad 
\la \Sigma \ra = v_D, \quad \la \Pi \ra=0
\ea
where $v_D= \sqrt{\lm_6} \et$.
We expand the fields $\ph_3$, $\ch$, $\Sigma$ and $ \Pi$ around their VEV,
\ba
\ph_3=v_{EW}+\hat{\ph}, \quad \ch=\et+\hat{\ch}, \quad \Sigma=v_D+ \sigma, 
\quad \Pi = \pi.
\ea 

From the scalar potential, \eqref{eq:S_EW_d}, and dark sector potential, 
\eqref{eq:Sigmamodel}, 
we obtain mixing between the fields $\hat{\ph}$, $\sigma$ and $\hat{\ch}$, 
having the following squared mass matrix,
\ba
\mc M^2=\left(
\begin{array}{ccc}
2 \lambda_1 \lambda_2 & 0 & -2\lambda_1 \lambda_2^{3/2} \\
 0 & 2\lambda_5\lambda_6 & -2\lambda_5 \lambda_6^{3/2} \\
 -2\lambda_1 \lambda_2^{3/2}  & -2\lambda_5 \lambda_6^{3/2} & 2(\lambda_1 \lambda_2^2+\lambda_5 \lambda_6^2) + \frac{m^2}{\eta^2} \\
\end{array}
\right) \et^2
\label{eq:massmatrix}
\ea
where $m^2=2\lambda\eta^2$ is the mass term generated by the 
contributions due to scale anomaly. 
The eigenvalues of this mass matrix gives us three physical scalar particles, which we denote as $\ch_1, \ch_2$ and $\ch_3$. We identify $\chi_1$ as the 125 GeV Higgs Boson. The mass matrix can be diagonalized  by an orthogonal matrix $R$ with three Euler angles ($\alpha_1,\alpha_2,\alpha_3$)

\begin{equation}
\\ R(\alpha_1,\alpha_2,\alpha_3)=\left(\begin{array}{ccc}
c_{\alpha_1}c_{\alpha_2} & -s_{\alpha_1}c_{\alpha_2} & s_{\alpha_2} \\
-c_{\alpha_1}s_{\alpha_2}s_{\alpha_3}+s_{\alpha_1}c_{\alpha_3} & c_{\alpha_1}c_{\alpha_3}+s_{\alpha_1}s_{\alpha_2}s_{\alpha_3} & c_{\alpha_2}s_{\alpha_3}\\
-c_{\alpha_1}s_{\alpha_2}c_{\alpha_3}-s_{\alpha_1}s_{\alpha_3} & -c_{\alpha_1}s_{\alpha_3}+s_{\alpha_1}s_{\alpha_2}c_{\alpha_3}& c_{\alpha_2}c_{\alpha_3}\\
\end{array}\right)
\end{equation}
and the mass eigenstates are obtained such that 
\begin{equation}
\left(
\begin{array}{c}
\phi\\
\sigma\\
\chi
\end{array}
\right)=
R
\left(
\begin{array}{c}
\chi_1\\
\chi_2\\
\chi_3
\end{array}
\right)
\end{equation}

The parameters of the model $(\lambda_1,\lambda_2,\lambda_5,\lambda_6,\lambda_7,m^2)$,
 written in terms of the physical masses, Euler angles (or the mixing angles) and the VEVs
are
\begin{eqnarray}
\lambda_1&=&\frac{(M_{\chi_1}^2 R_{11}^2 + M_{\chi_2}^2 R_{12}^2 + M_{\chi_3}^2 R_{13}^2)^3}{2 \eta^2(M_{\chi_1}^2 R_{11} R_{31} + M_{\chi_2}^2 R_{12} R_{32} +M_{\chi_3}^2 R_{13} R_{33})^2}\nonumber \\[18pt]
\lambda_2&=&\frac{(M_{\chi_1}^2 R_{11} R_{31} + M_{\chi_2}^2 R_{12} R_{32} + M_{\chi_3}^2 R_{13} R_{33})^2}{(M_{\chi_1}^2 R_{11}^2 + M_{\chi_2}^2 R_{12}^2 + M_{\chi_3}^2 R_{13}^2)^2}\nonumber \\[18pt] 
\lambda_5&=&\frac{(M_{\chi_1}^2 R_{21}^2 + M_{\chi_2}^2 R_{22}^2 + M_{\chi_3}^2 R_{23}^2)^3}{2 \eta^2(M_{\chi_1}^2 R_{21} R_{31} + M_{\chi_2}^2 R_{22} R_{32} +M_{\chi_3}^2 R_{23} R_{33})^2}\nonumber \\[18pt] 
\lambda_6&=&\frac{(M_{\chi_1}^2 R_{21} R_{31} + M_{\chi_2}^2 R_{22} R_{32} + M_{\chi_3}^2 R_{23} R_{33})^2}{(M_{\chi_1}^2 R_{21}^2 + M_{\chi_2}^2 R_{22}^2 + M_{\chi_3}^2 R_{23}^2)^2}\nonumber \\[18pt]\nonumber
\end{eqnarray}
\begin{eqnarray}  
\lambda_7&=&\frac{M_{\pi}^2}{\eta^2} \nonumber \\[18pt]
m^2&=&M_{\chi_1}^2 R_{31}^2 + M_{\chi_2}^2 R_{32}^2 + M_{\chi_3}^2 R_{33}^2\nonumber \\
&& - (M_{\chi_1}^2 R_{11}^2 + M_{\chi_2}^2 R_{12}^2 + M_{\chi_3}^2 R_{13}^2)\lambda_2 \nonumber \\
&& - (M_{\chi_1}^2 R_{21}^2 + M_{\chi_2}^2 R_{22}^2 + M_{\chi_3}^2 R_{23}^2)\lambda_6\\ \nonumber
\end{eqnarray}

The VEVs are related by the equations

\begin{eqnarray}
v_{D}^2=\lambda_6\eta^2 \hspace{5mm} \text{and} \hspace{5mm}  v_{EW}^2=\lambda_2\eta^2 
\end{eqnarray}  
The mass of the physical scalars ($\chi_1,\chi_2,\chi_3$) are related by the relation
\begin{equation}
M_{\chi_1}^2R_{11}R_{21} + M_{\chi_2}^2R_{12}R_{22} + M_{\chi_3}^2R_{13}R_{23}=0
\end{equation} 

The independent parameters of the model are $\alpha_1,\alpha_2,\alpha_3,M_{\chi_1},M_{\chi_2},M_{\pi}$ and $v_{EW}$ and we set $M_{\chi_1}$ and $v_{EW}$ to be 125 GeV and 246 GeV respectively. We take $\chi_1$ to be the lightest observed Higgs and assume $\chi_2$ and $\chi_3$ heavier than $\chi_1$.  
 Due to scale anomaly $\chi_3$ acquires a mass which we assume to be larger than
the mass of the Higgs boson. The
Feynman rules and relevant parameters in the model are implemented with FeynRules \cite{Alloul:2013bka}.

\section{Phenomenology: Freeze out Scenario}
\label{sec:phenomenology}
 In this section, we discuss the phenomenological implications of 
our model assuming that the relic density of DM is obtained
in the freeze out scenario. We investigate the constraints from the Higgs sector such as the branching fraction of the Higgs to the visible sector and the upper bound on the Higgs decay to the invisible sector.  
 We also explore the SI scattering cross-section of 
pions with the nucleons and the $\gamma$-ray constraints termed as direct and indirect detection of DM respectively. For simplicity we considered the mixing angles in the region $[0,\frac{\pi}{2}]$. For the dark pion mass we consider $M_{\pi}\in [50-400]$ GeV and, 
as we shall see, some of the allowed parameter space lies 
 close to the resonance
of $\chi_1$, $\chi_2$ and $\chi_3$.
 From the perspective of phenomenology the dark sector neutral scalars $\chi_2$ and $\chi_3$ share the same Higgs like properties with mass greater than 125 GeV. Due to
 the presence of additional two Higgs like scalars, the analysis of our 
model has some similarities with Higgs portal DM \cite{McDonald:1993ex,Burgess:2000yq,Arcadi:2019lka,Athron:2018hpc,Athron:2017kgt}.  

\subsection{Collider constraints}
\label{sec:collider}
\begin{figure}[ht]
\begin{subfigure}{.5\textwidth}
  \centering
  % include first image
  \includegraphics[width=1.1\linewidth]{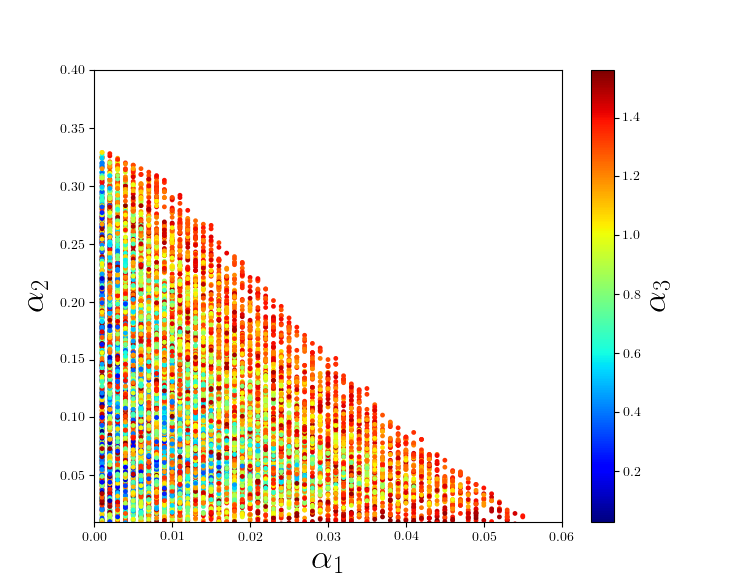}  
  \caption{}
  \label{HS1}
\end{subfigure}
\begin{subfigure}{.5\textwidth}
  \centering
  % include second image
  \includegraphics[width=1.1\linewidth]{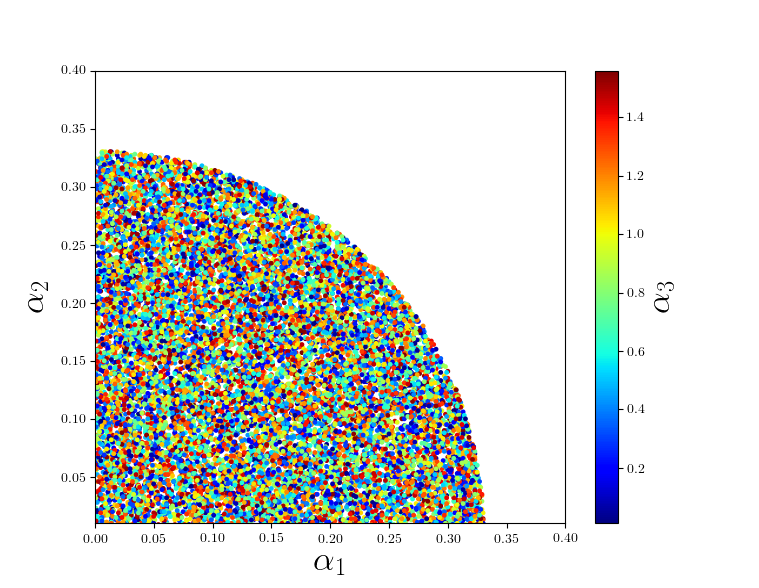}  
  \caption{}
  \label{HS2}
\end{subfigure}
\caption{The $1\sigma$ allowed region of mixing angles are shown. Fig (a) shows the allowed region for $M_\pi = 60$ GeV and $M_{\chi_2} = 500$ GeV. Fig (b) shows the allowed region for $M_\pi \geq M_{\chi_1}/2$.}
\label{decaywidth}
\end{figure}

 The deviations from the SM can be implemented by introducing the coupling modifiers $(\kappa's)$,
\begin{equation}
g_{\chi_1f\bar{f}}=\kappa_fg^{\text{SM}}_{hf\bar{f}} \hspace{5mm} \text{and}\hspace{5mm} g_{\chi_1VV}=\kappa_Vg^{\text{SM}}_{hVV} \ ,
\end{equation}
such that $\kappa_f=1$ and $\kappa_V=1$ in SM. Here $g^{\text{SM}}_{hf\bar{f}}$
and $g^{\text{SM}}_{hVV}$ represent the SM Higgs couplings to fermions and vector bosons respectively and $g_{\chi_1f\bar{f}}$ and $g_{\chi_1VV}$ represent the
corresponding couplings in our model. The coupling modifiers are given 
by $\kappa_{f}=\kappa_{V}=\cos{\alpha_1}\cos{\alpha_2}$.
The constraints on these parameters from the Higgs coupling 
measurements
are given in \cite{ATLAS:2018doi,Sirunyan:2018koj}. Using the latest beta version of HiggsSignals \cite{Bechtle:2013xfa} we have imposed the constraints on the mixing angles as shown in Figs. \ref{HS1} and \ref{HS2}. For $M_{\pi} \geq M_{\chi_1}/2$ the allowed region in the $\alpha_1-\alpha_2$ plane is a radius of approximately 0.32 radians, leaving  $\alpha_3$ unconstrained. The invisible decay of Higgs opens up when $M_{\pi}$ goes below $M_{\chi_1}/2$, we can express this branching fraction as
\begin{equation}
Br^{inv}=\frac{\Gamma^{\chi_{1}\rightarrow\pi\pi}}{\Gamma^{\chi_{1}\rightarrow\pi\pi}+ R_{11}^2\Gamma^{SM}}
\end{equation}
where $\Gamma^{SM}=4.07$ MeV, is the SM Higgs decay width. The observed upper limit on Higgs invisible branching fraction is 0.24 \cite{Khachatryan:2016whc,Aaboud:2019rtt}. The decay width of $\chi_{1}$ 
to dark pions is given by 
\begin{equation}
\Gamma^{\chi_1\rightarrow\pi\pi}=\frac{g_{\chi_1\pi\pi}^2}{32M_{\chi_1}\pi} \sqrt{1-\frac{ 4M_{\pi}^2}{M_{\chi_1}^2}}
\end{equation}
 We see that the mixing angles are much contrained when the Higgs invisible decay to dark pions is open.

\subsection{Relic density calculation}
\label{sec:relic}

%\begin{figure}[h]
 % \centering
 % \begin{minipage}[b]{0.35\textwidth}
 %   \includegraphics[width=\textwidth]{relic_fig_1.pdf}
 % \end{minipage}
 % \hfill
 % \begin{minipage}[b]{0.37\textwidth}
 %   \includegraphics[width=\textwidth]{relic_fig_2.pdf}
 % \end{minipage}
 % \hfill
 % \begin{minipage}[b]{0.37\textwidth}
 %   \includegraphics[width=\textwidth]{relic_fig_3.pdf}
 %% \end{minipage}
  %\hfill
  %\begin{minipage}[b]{0.37\textwidth}
  %  \includegraphics[width=\textwidth]{relic_fig_4.pdf}
  %\end{minipage}
  % \caption{Diagrams contributing to the $\langle\sigma v\rangle$ calculation of relic density. In our notation the light fermions $f=b,s,c,u,d,\tau,\mu,e$, the vector bosons $V=W^{\pm},Z$ and $\phi=a_{11}\chi_1 + a_{12}\chi_2 + a_{13}\chi_3$. As the dark pion mass is assumed to be $40$ GeV throughout our analysis the most dominant contribution comes from the annihilation into light fermions }\label{relic_plot}
  %\end{figure}

\begin{figure}[ht]
\begin{subfigure}{.5\textwidth}
  \centering
  % include first image
  \includegraphics[width=1\linewidth]{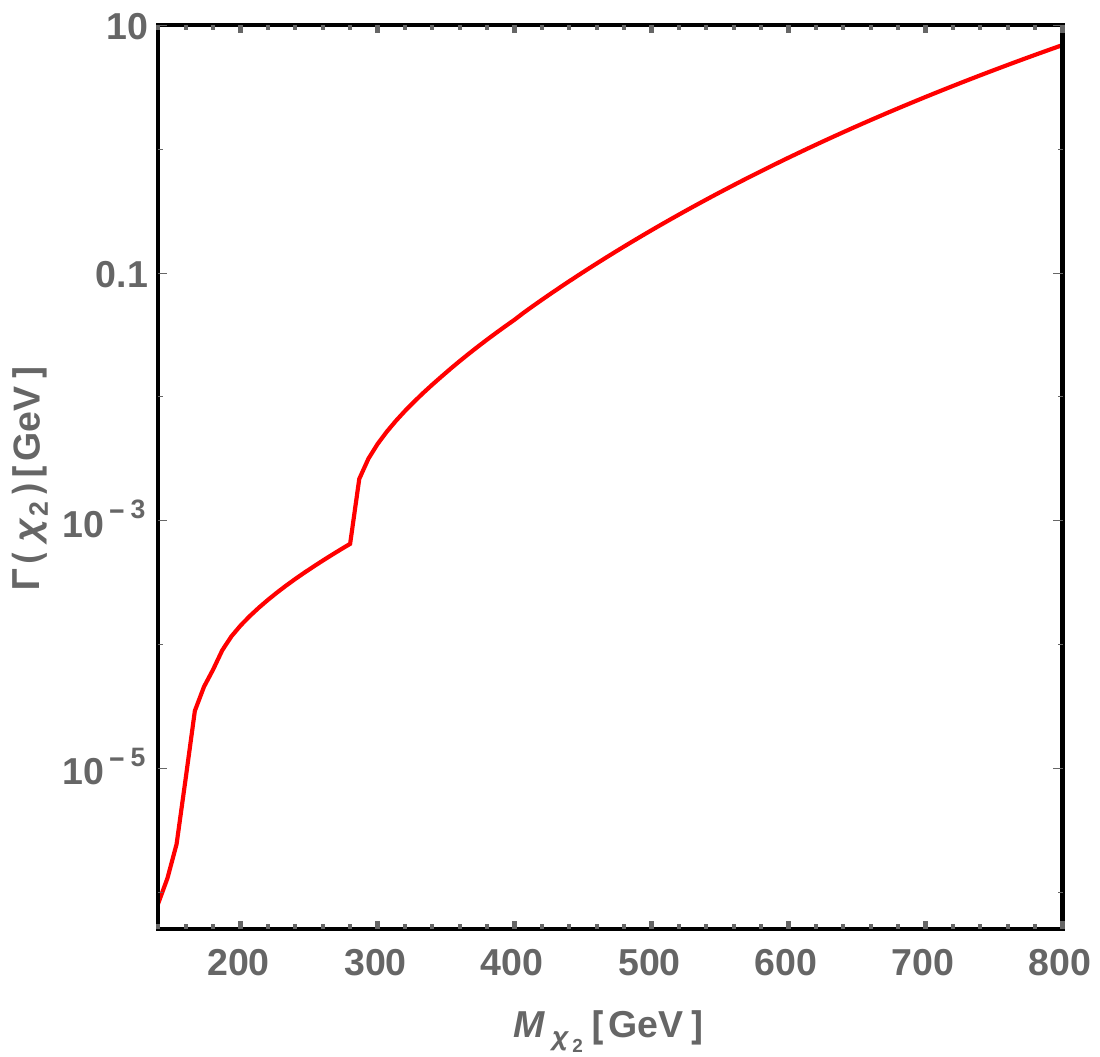}  
  \caption{}
  \label{h2decay1}
\end{subfigure}
\begin{subfigure}{.5\textwidth}
  \centering
  % include second image
  \includegraphics[width=1\linewidth]{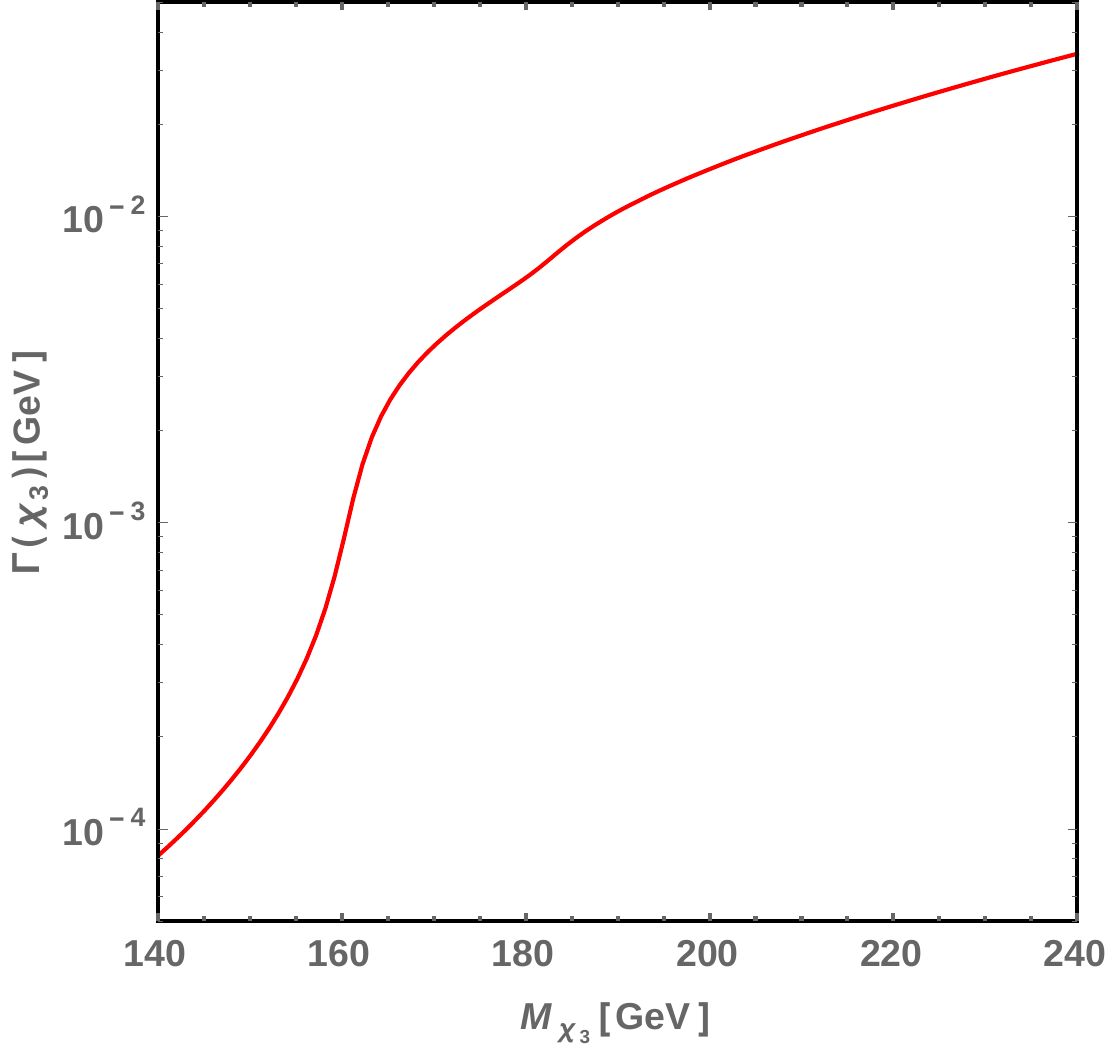}  
  \caption{}
  \label{h3decay1}
\end{subfigure}
\caption{The decay width of $\chi_2$ and $\chi_3$ for $\alpha_1=0.01 $ radians, $\alpha_2=0.1 $ radians, $\alpha_3=1.0 $ radians and dark pion mass of 200 GeV. The decay width is sufficiently large compared to $H_0$.}
\label{decaywidth1}
\end{figure}

In this section we calculate the relic density of dark pions within the framework of freeze out scenario. As explained earlier, the dark sector of our model 
contains dark pions and scalar particles ($\chi_2$
and $\chi_3$). The scalar particles have masses of the order of the dark strong interaction scale. These   
decay into dark pions early
in the Universe, leaving pions as the only DM candidate. The total decay width of $\chi_2$
and $\chi_3$ as a function of their masses is computed
 using HDECAY \cite{Djouadi:1997yw} and CalcHEP \cite{Belyaev:2012qa} and 
 plotted in Fig. \ref{decaywidth1}.
We require that the lifetime of $\chi_2$
and $\chi_3$  be smaller than the lifetime of the Universe ($H_0^{-1}$ ), 
i.e. $\Gamma_{\chi_2}$ and $\Gamma_{\chi_3}$  be greater than $H_0$, which is satisfied in this case, as shown in Fig. \ref{decaywidth1} . We solve the Boltzmann equation to find the freeze out temperature and the comoving number density of dark 
pions. The Boltzmann equation \cite{Gondolo:1990dk,Edsjo:1997bg} for comoving number density of dark pions can be written as  

\begin{eqnarray}
\frac{dY}{dx}=-\Big( \frac{45 ~G}{\pi} \Big)^{-1/2} \frac{g_{\star}^{1/2}M_{\pi}}{x^2}\langle \sigma v \rangle
(Y^2-Y_{eq}^2)
\label{eq:bolt}
\end{eqnarray}
where $Y=\frac{n_{\pi}}{s}$, $x=\frac{M_{\pi}}{T}$, $s$ is the entropy density and $g_{\star}$ is the number of degrees of freedom. This is given by  
  \begin{eqnarray}
g_{\star}^{1/2}=\frac{h_s(T)}{g_{\rho}(T)^{1/2}}\Big(1+\frac{1}{3}\frac{T}{h_s(T)}\frac{dh_s(T)}{dT} \Big)
\end{eqnarray}
where $g_s(T)$ and $g_{\rho}(T)$ are the effective degrees of freedom related to entropy and energy density respectively. Thermal average cross section $\langle \sigma v\rangle$ is given by 
  \begin{eqnarray}
\langle \sigma v\rangle = \frac{1}{8 M_{\pi}^4 T K_2^2(x)} \int_{4 M_{\pi}^2}^{\infty}  ds~ \sigma (s-4 M_{\pi}^2)\sqrt{s}~K_1\left(\frac{\sqrt{s}}{T}\right)
\end{eqnarray}
where $K_1$ is the Bessel function. Integrating Eq. \eqref{eq:bolt} from 
$x=0$ to $\frac{M_{\pi}}{T_0}$, where $T_0= 2.72 K$ is the temperature of the
  cosmic microwave background radiation, we get the current value of $Y$, i.e. $Y_0$, and hence
the relic density. The present relic density is given by 
\begin{eqnarray}
\Omega_{\pi} h^2= 2.755 \times 10^8\left(\frac{M_{\pi}}{GeV}\right) Y_0
\end{eqnarray}
 
We use the MicrOMEGAs package \cite{Belanger:2014vza} to
calculate the relic density of dark pion.
Here $\pi ~ \pi \rightarrow \chi_{1,2,3} \rightarrow SM ~SM $ processes contribute to the total 
thermally averaged cross section. The value of this cross section at the decoupling temperature determines the relic density of the dark pion and the s-channel processes involved in the cross section might encounter poles. 
We find that the process 
$\pi ~ \pi \rightarrow \chi_{1,2,3} \rightarrow SM ~SM $ also gets resonant contributions
when the mass $M_\pi$ is close to half the mass of the scalars ($\chi_1, \chi_2$ and $\chi_3$). 
As a result of the resonance there is a peak in $\langle \sigma v \rangle$ 
characterized by the parameter $\epsilon=\Gamma_{\text{res}}/M_{\text{res}}$,
 where $M_{\text{res}}$ is the mass of the resonance particle $\chi_1, \chi_2$ and $\chi_3$. This leads to an enhancement in comparison 
to the non resonant cross section\footnote{The most common approach of Taylor expansion of $\langle\sigma v\rangle=a +  b v^2$ and then the substitution of $v^2=6/x$ does not hold at resonance as shown in Ref.\cite{Griest:1990kh}. The numerical result of $\langle \sigma v \rangle $ differs sharply from the approximate result obtained from Taylor expansion and the disagreement is sharper for smaller $\epsilon$. }. Enhancement of annihilation cross section at resonance has been discussed in great details  in Refs. \cite{Griest:1990kh,Ibe:2008ye,Guo:2009aj,Duch:2017nbe}. The enhancement of the cross section at resonance leads to a sudden drop of relic density.

\subsection{Direct Detection and Indirect Detection Constraints }
\label{sec:direct}
Direct detection experiments impose strong limits on the interaction 
of dark matter particles with nucleons. In these experiments
dark matter particles are scattered elastically with the nucleons present in the detectors and the recoil energy is detected. Experiments can be sensitive to both nuclear spin-independent (SI) interactions and spin-dependent (SD) interactions but current experiments are more sensitive to SI interactions. The results obtained from 
LUX \cite{Akerib:2016vxi,Akerib:2015rjg} and PandaX-II \cite{Cui:2017nnn,Tan:2016zwf} put upper bounds on SI scattering cross-section. Latest result obtained from XENON1T \cite{Aprile:2018dbl} experiment being more restrictive excludes a large parameter space in many DM models as the upper limit on the SI cross-sections 
gets pushed to an order of $10^{-47}$ cm$^2$ for DM mass of around 50 GeV at $90\%$ confidence limit. These results impose a very strong constraint,
 particularly close to the Higgs resonance ($M_\pi\sim 60$ GeV).  
   %\cite{PhysRevLett.108.259002,Aalseth:2012if,Bernabei:2010mq, Agnese:2013rvf,Cushman:2013zza,Angle:2011th,Aprile:2011hi}.

%\begin{figure}[h!]
 % \centering
 % \begin{minipage}[b]{0.30\textwidth}
 %   \includegraphics[width=\textwidth]{dd.pdf}
 % \end{minipage}
 % \caption{The scattering of dark pions with nucleons by exchange of $\chi_1$.}
 % \hfill
%\label{fig:piNscattering}
 % \end{figure}

\begin{figure}{}
\begin{subfigure}{.5\textwidth}
  \centering
  % include first image
  \includegraphics[width=.9\linewidth]{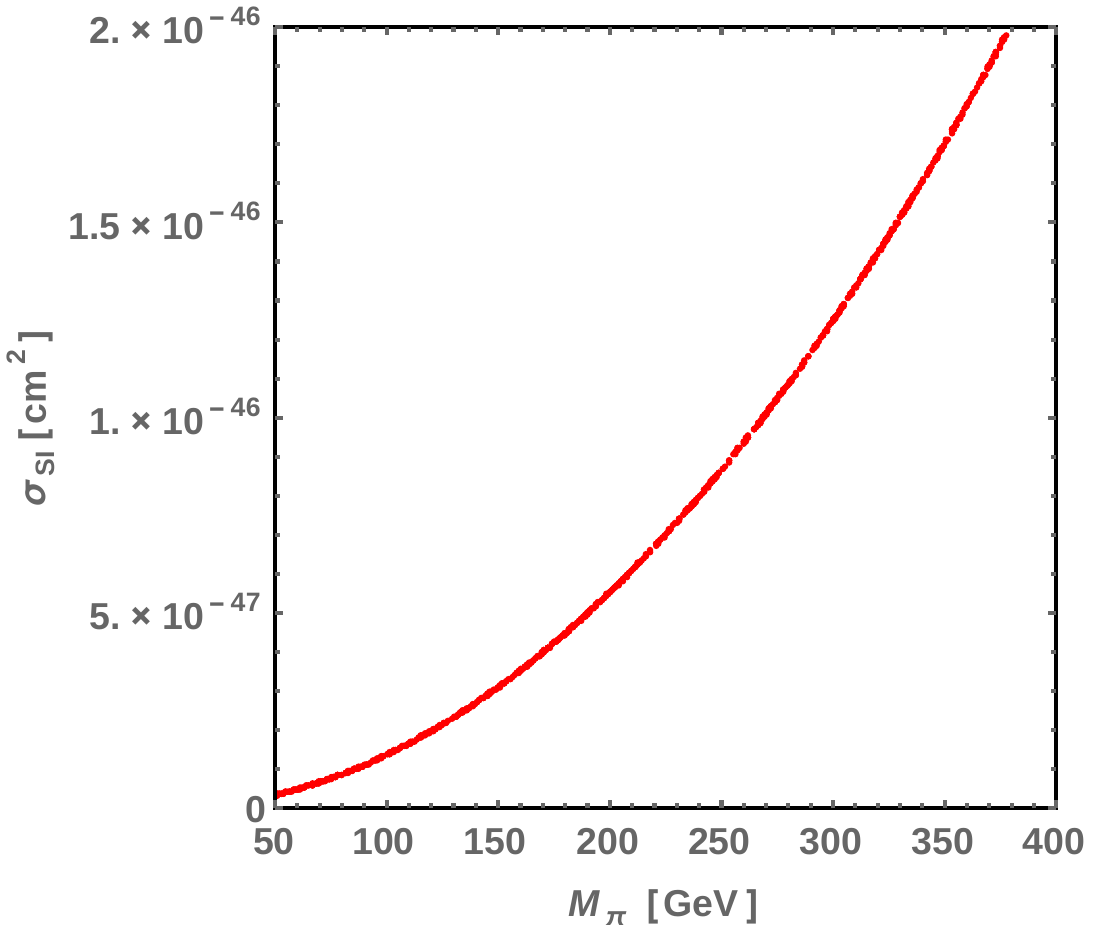}  
  \caption{}
  \label{dd}
\end{subfigure}
\begin{subfigure}{.5\textwidth}
  \centering
  % include first image
  \includegraphics[width=0.8\linewidth]{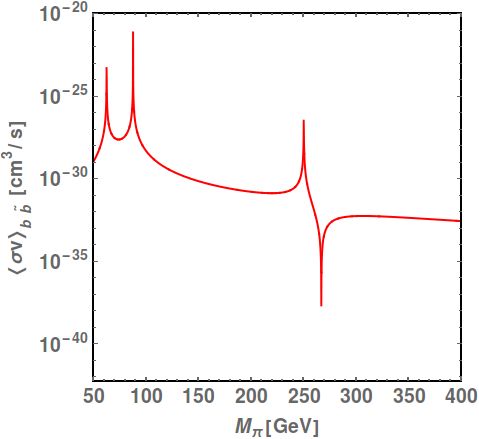}  
  \caption{}
  \label{bb}
\end{subfigure}
\begin{subfigure}{.6\textwidth}
  \centering
  % include first image
  \includegraphics[width=0.7\linewidth]{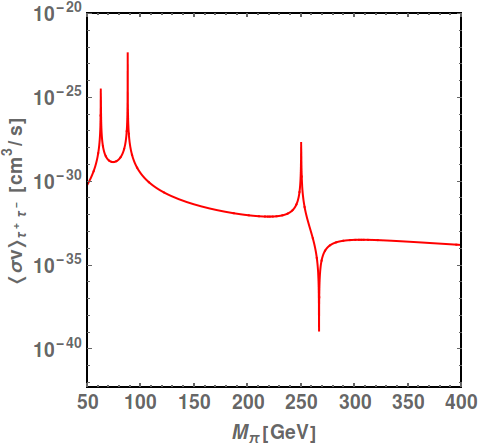}  
  \caption{}
  \label{tata}
\end{subfigure}
\begin{subfigure}{.5\textwidth}
  \centering
  % include first image
  \includegraphics[width=0.8\linewidth]{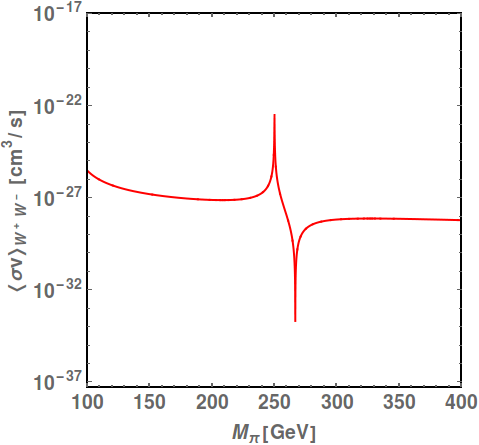}  
  \caption{}
  \label{ww}
\end{subfigure}
\begin{subfigure}{.5\textwidth}
  \centering
  % include first image
  \includegraphics[width=0.8\linewidth]{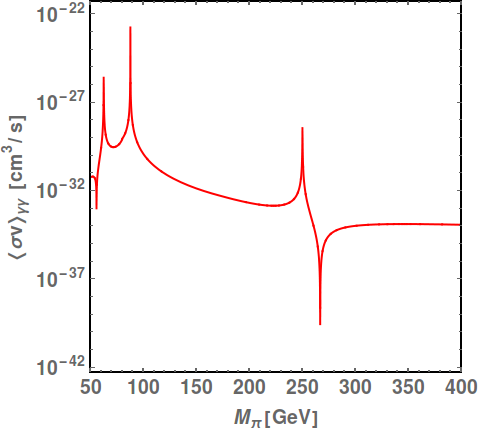}  
  \caption{}
  \label{gam_gam}
\end{subfigure}
\begin{subfigure}{.5\textwidth}
  \centering
  % include second image
  \includegraphics[width=0.8\linewidth]{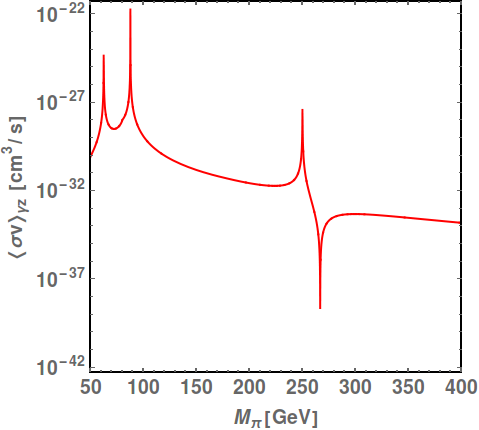}  
  \caption{}
  \label{gam_z}
\end{subfigure}
\caption{ (a) The dark pion-nucleon scattering cross-section is shown for dark 
pion mass in the range $[50-400]$ GeV for the mixing angles $\alpha_1=0.01$ radians $\alpha_2=0.1$ radians, $\alpha_3=1.0$ radians and $M_{\chi_2}=500$ GeV. 
For the same set of parameters, Figs. (b), (c), (d), (e)
and (f) show the cross sections for dark pion annihilation into $b\bar b$, $\tau^+ \tau^-$, $W^+W^-$, 
$\gamma\gamma$ and $\gamma Z$ respectively.
We see from these plots that the $\gamma$-ray constraints from dSphs
and galactic centre are automatically satisfied and are much weaker than the direct detection
constraints. The peaks appear due to the Breit Wigner resonance of $\chi_1$, $\chi_2$, $\chi_3$.
 The troughs in these plots arise since some coupling factors cross zero at the corresponding values of the pion mass.  }
\label{cross-section_plot}
\end{figure}

 The expression for SI elastic scattering cross section between dark pion and nucleon (N) through the exchange of $\chi_1,\chi_2$ and $\chi_3$ is given by
 
\begin{eqnarray}
\sigma^{\pi- N}_{SI}=\frac{\mu^2_{\pi N} m_N^2 f^2}{\pi M_\pi^2 v_{EW}^2}\Big(\frac{R_{11}g_{\chi_1\pi\pi}}{M_{\chi_1^2}} + \frac{R_{12}g_{\chi_2\pi\pi}}{M_{\chi_2^2}} + \frac{R_{13}g_{\chi_3\pi\pi}}{M_{\chi_3^2}}\Big)^2
\label{sigma_nucleon} 
\end{eqnarray}
where $f\sim$ 0.3 \cite{Cline:2013gha} is the usual nucleonic matrix element and 
$\mu_{\pi N}$ is the 
reduced mass of dark pion and nucleon. We compute the DM-nucleon SI cross-section using MicrOMEGAs. The SI cross-section of dark pions with the nucleons is 
plotted in Fig. \ref{cross-section_plot} for pion $\in[50,400]$ GeV.
 % The upper limit on the spin independent cross section from LUX direct detection observation is $7.6 \times 10 ^{-46}\ {\rm cm}^2$ for a dark matter of mass range 
%10 GeV-100 GeV.
%  The left panel of Fig.\ref{region_plot} shows the allowed region of $v_D$ and $\eta$ for 40 GeV pion. We consider two choices of the parameter $\lambda_5=0.06$ (light grey) and $\lambda_5=0.1$ (dark grey) respectively by using the constraints on the spin-independent cross-section of dark matter with  nucleons. 
%We see that the parameter values used to fit the relic density and
%the gamma ray signal are allowed by direct detection experiments by lowering the parameter $\lambda_5$ below 0.1.

  We next turn to the indirect detection of DM where we primarily look into the gamma ray signals from the dSphs and monoenergetic spectrum from the galactic centre. The diffused gamma ray search studied by the combined analysis of $Fermi$-LAT and MAGIC observations put limits on the DM scattering cross-section to $b\bar{b}$, $W^+W^-$, $\tau^+\tau^-$ and $\mu^+\mu^-$.  The cross sections are given by \\  
 \begin{eqnarray}
\langle\sigma v\rangle_{f\overline{f}} &=& \frac{N_c}{4 \pi}\Big(\frac{m_f}{v_{\text{EW}}}\Big)^2 \left| \frac{R_{11}g_{\chi_1 \pi \pi}}{s-M_{\chi_1}^2+i M_{\chi_1}\Gamma_{\chi_1}} +\frac{R_{12}g_{\chi_2 \pi \pi}}{s-M_{\chi_2}^2+i M_{\chi_2}\Gamma_{\chi_2}} +\frac{R_{13}g_{\chi_3 \pi \pi}}{s-M_{\chi_3}^2+iM_{\chi_3}\Gamma_{\chi_3}} \right|^2 \nonumber \\
&&\times\Big ( 1- \frac{4 m_f^2}{s}\Big)^{3/2} \
\label{sigmavbb}  
\end{eqnarray} 
where $N_c$ is the color charge of the fermion $f$.

\begin{eqnarray}
\langle\sigma v\rangle_{W^+ W^-}&=&\frac{s}{2\pi v_\text{EW}^2} \left| \frac{R_{11}g_{\chi_1 \pi \pi}}{s-M_{\chi_1}^2+i M_{\chi_1}\Gamma_{\chi_1}} +\frac{R_{12}g_{\chi_2 \pi \pi}}{s-M_{\chi_2}^2+i M_{\chi_2}\Gamma_{\chi_2}} +\frac{R_{13}g_{\chi_3 \pi \pi}}{s-M_{\chi_3}^2+iM_{\chi_3}\Gamma_{\chi_3}} \right|^2 \nonumber \\ 
&&\times \sqrt{1-\frac{4 M_W^2}{s}} \Big(1-\frac{4M_W^2}{s} + \frac{3}{4}\Big(\frac{4 M_W^2}{s}\Big)^2\Big)
\end{eqnarray}
 
The dark pion annihilation to $\gamma\gamma$ and $Z\gamma$ is important
for detection of monochromatic gamma ray signal from the galactic center. The 
cross-sections for these processes are loop suppressed and are given by 
\cite{Feng:2014vea}
\begin{eqnarray}
\langle\sigma v\rangle_{\gamma\gamma/\gamma Z}&=&\frac{8}{\sqrt{s}}\left|\frac{R_{11}g_{\chi_1\pi\pi}}{(s-M_{\chi_1}^2)+iM_{\chi_1}\Gamma_{\chi_1}} + \frac{R_{12}g_{\chi_2\pi\pi}}{(s-M_{\chi_2}^2)+iM_{\chi_2}\Gamma_{\chi_2}} + \frac{R_{13}g_{\chi_3\pi\pi}}{(s-M_{\chi_3}^2)+iM_{\chi_3}\Gamma_{\chi_3}}\right|^2 \nonumber \\
&& \times\Gamma_{\gamma\gamma/\gamma Z}(s)
\label{sigmavgamgam}
\end{eqnarray}  
where $\Gamma_{\gamma\gamma/\gamma Z}(s)$  
is computed by replacing the scalar mass square in the Higgs 
decay rate into
 $\gamma\gamma/\gamma Z$ with $s$ \cite{Gunion:1989we,Djouadi:2005gi,Dittmaier:2011ti}.
  In the non-relativistic limit we can consider the pions to be almost at rest (which is a very good approximation) and in this limit $s\sim 4 M_{\pi}^2$. The results are consistent with MicrOMEGAs and cross-sections are shown in Fig. \ref{cross-section_plot}. The DM distribution in the dSphs are parameterized following a Navarro-Frenk-White (NFW) profile \cite{Navarro:1995iw}, whereas for the galactic DM halo we consider the Einasto \cite{1968PTarO..36..414E} profile as it is more restrictive compared to the NFW profile.

\subsection{Dark Matter Self Interaction Constraints}
 The observed central densities of dark matter halos of a wide range
  of the astrophysical objects from dwarf galaxies to galaxy clusters have lesser
 density compared
  to the prediction from the collisionless cold dark matter N-body simulations \cite{Flores:1994gz}. This mass
  deficit anomaly could be resolved if the cold dark matter particles
 undergo elastic scattering among
  one another. The self-interacting dark matter (SIDM) leads to 
exchange of heat energy between
  inner and outer halos and leads to a lower density of the inner halos \cite{Peter:2012jh,Rocha:2012jg,Elbert:2014bma,Dave:2000ar}. Hence, the self-interaction of dark matter is able solve this ‘too-big-to-fail’ problem \cite{Garrison,Boylan} and also the  ‘cusp vs.  core’ problem \cite{Flores:1994gz, Navarro:1996gj, Oh_2011, Walker_2011}. Astrophysical observations suggest the following value for DM-DM scattering 
cross section ($\sigma_{DM}$) \cite{Kahlhoefer:2015vua}
\begin{equation}
{\sigma_{DM}\over M_{\text{DM}}} \sim 1.5 \ {\rm cm}^2\ {\rm g}^{-1}
\label{eq:sigmaoverm}
\end{equation}
where $M_{\text{DM}}$ is the mass of the dark matter. 
%Due to the large uncertainty we use the round figure of 1 cm$^2$/g. 
This estimate is subject 
to some uncertainties. There also exists an upper bound  \cite{Harvey:2015hha,Kahlhoefer:2015vua,Randall:2007ph,Markevitch:2003at,Peter:2012jh,Rocha:2012jg}
\begin{equation}
{\sigma_{DM}\over M_{\text{DM}}} \lesssim 1 \ {\rm cm}^2\ {\rm g}^{-1}\ ,
\label{eq:sigmaoverm1}
\end{equation}
obtained from different astrophysical observations. 
These considerations 
suggest that the dark pion - dark pion ($\pi\, \pi\rightarrow \pi\, \pi$)
 scattering cross section $\sigma_\pi$ should satisfy $\sigma_{DM}/M_{DM} \in [4.7-7.0]\times 10^3$ GeV$^{-3}$ \cite{Campbell:2015fra}.

\section{Phenomenology: Freeze in Scenario} 
\label{sec:freezein}

In this section we examine the implications of the model assuming the 
freeze in scenario \cite{Hall:2009bx,Babu:2014pxa,Belanger:2018mqt}. 
In this scenario, DM particles are never in equilibrium with the cosmic plasma 
and at very high temperature their density is zero. 
The production of dark pions happens  
through $SM ~SM\rightarrow \pi \pi$ and $\chi_1 \rightarrow \pi \pi$ 
processes. The freeze in scenario within the 
Higgs portal DM models successfully explains the astrophysical constraints coming from DM self interaction \cite{Campbell:2015fra,Bernal:2017kxu,Kang:2015aqa}.
As we shall see this is also true in our conformal model.
We will use self interacting DM scattering constraints to put limits on the parameter space.

In the case of freeze in we find that we can satisfy all the constraints
provided we choose both the parameters $\lambda_2$ and $\lambda_6$ 
relatively small. We shall assume $\lambda_2<<\lambda_6<<1$ which
 leads to $\eta>> v_D>> v_{EW}$.  
We also need to choose $\lambda_7<<1$ due to the low value of pion mass in this 
scenario. The low value of $\lambda_7$ does not require any fine tuning since this parameter
corresponds to a symmetry breaking term.  
Furthermore the small values of $\lambda_2$ and $\lambda_6$ also do not require fine tuning
since these do not acquire large contributions at loop orders.
In the limit $\lambda_2<<\lambda_6<<1$ we can diagonalize the mass matrix, Eq. \ref{eq:massmatrix},
 perturbatively and relate the particles 
 $\hat \phi$, $\sigma$ and $\hat\chi$ to the
physical particles $\chi_1$, $\chi_2$ and $\chi_3$. 
At leading order, we obtain
\ba
\hat\phi &\approx & {1\over \sqrt{1+a_3^2}}\chi_1 + {c_1\over \sqrt{1+c_1^2+c_2^2}}\chi_3\nonumber\\
\sigma &\approx & {1\over \sqrt{1+b_3^2}}\chi_2 + {c_2\over \sqrt{1+c_1^2+c_2^2}}  \chi_3\nonumber\\
\hat\chi &\approx &  {1\over \sqrt{1+c_1^2+c_2^2}} \chi_3 + {a_3\over \sqrt{1+a_3^2}}\chi_1 + {b_3\over \sqrt{1+b_3^2}}   \chi_2
\ea
where $c_1$, $c_2$, $a_3$ and $b_3$ are all small compared to unity and given by
\ba
c_1 &=& {M_{\phi\chi}\over M_\chi^2 - M_\phi^2}\,,\nonumber\\
c_2 &=& {M_{\sigma\chi}\over M_\chi^2 - M_\sigma^2}\,, \nonumber\\
a_3 &=& {M_{\phi\chi}\over M_\phi^2 - M_\chi^2}\,, \nonumber\\
b_3 &=& {M_{\sigma\chi}\over M_\sigma^2 - M_\chi^2}
\ea
and
\ba
M^2_\phi &=& 2\lambda_1\lambda_2 \eta^2 \approx M_{\chi_1}^2\,,\nonumber\\
M^2_\sigma &=& 2\lambda_5\lambda_6 \eta^2 \approx M_{\chi_2}^2\,,\nonumber\\
M^2_\chi &\approx& m^2 \approx M_{\chi_3}^2\,,\nonumber\\
M_{\phi\chi} &=& -2{\lambda_1\lambda_2^{3/2}}\ \eta^2\,, \nonumber\\
M_{\sigma\chi} &=& -2\lambda_5\lambda_6^{3/2} \eta^2\,.
\ea 
Here we have assumed that $2(\lambda_1\lambda_2^2 +\lambda_5\lambda_6^2)<< m^2/\eta^2$.
The masses of the three physical scalars $\chi_1$, $\chi_2$ and $\chi_3$ are 
$M_{\chi_1}$, $M_{\chi_2}$ and $M_{\chi_3}$ respectively and $M_\pi = 
\sqrt{\lambda_7}\, \eta$. 
It is clear that $ M_{\chi_2}>> M_{\chi_1}$. 
Furthermore we shall choose $m$ such that $M_{\chi_3}>> M_{\chi_2}$.

In the freeze in scenario the relic density of dark matter, i.e. dark pions, is given by \cite{Bernal:2018ins,Bernal:2017kxu}
\begin{equation}
{\Omega_{DM}h^2\over 0.12} = 5.3\times 10^{21} \lambda_{h\pi}^2 {M_\pi\over {\rm GeV}}
\label{eq:relicFreezein}
\end{equation}
where $\lambda_{h\pi}v_{EW}$ is the coupling of the Higgs to two dark pions and is given by
\begin{equation}
 \lambda_{h\pi} = {8\lambda_1\lambda_2\lambda_6\over \lambda_5\lambda_6^2 + 
m^2/(2\eta^2)}
\label{eq:couplingFreezein}
\end{equation}

In our model $\pi\, \pi\rightarrow \pi\, \pi$ scattering cross section
 gets contribution from the 
contact interaction $\Pi^4$ term as well as due to exchange of the three
physical scalars $\chi_1$, $\chi_2$ and $\chi_3$. The interaction of
dark pions with $\chi_1$ and $\chi_3$ is very small and hence we only need
to include the contributions due to $s$, $t$ and $u$ channel exchange
of $\chi_2$. At low energies these give contributions to the scattering
amplitude which are proportional to that obtained from the self interaction
$\Pi^4$. The final result is found to be 
\begin{equation}
{\sigma_{\pi} \over M_\pi} = {9\tilde\lambda_5^2\over 32\pi} {1\over M_\pi^3} 
\label{eq:sigmaovermpi}
\end{equation}
where $\tilde\lambda_5 = 3\lambda_5/4$.

\section{Results: Freeze out Scenario}

\label{sec:results}
\begin{figure}{}
	\begin{subfigure}{.55\textwidth}
		\centering
		% include first image
		\includegraphics[width=1.1\linewidth]{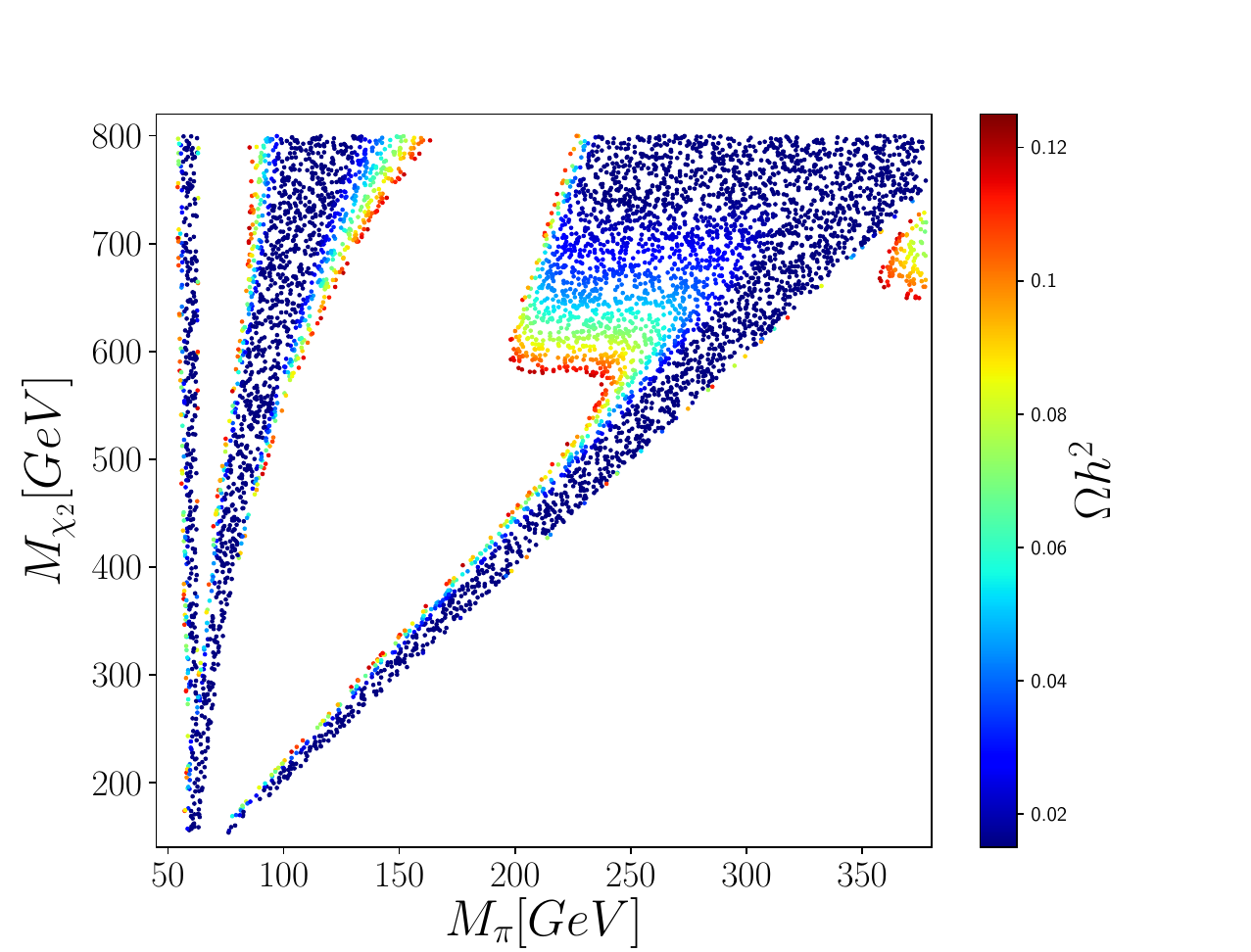}  
		\caption{}
		\label{fig:sub-4th}
	\end{subfigure}
	\begin{subfigure}{.5\textwidth}
		\centering
		% include second image
		\includegraphics[width=1.1\linewidth]{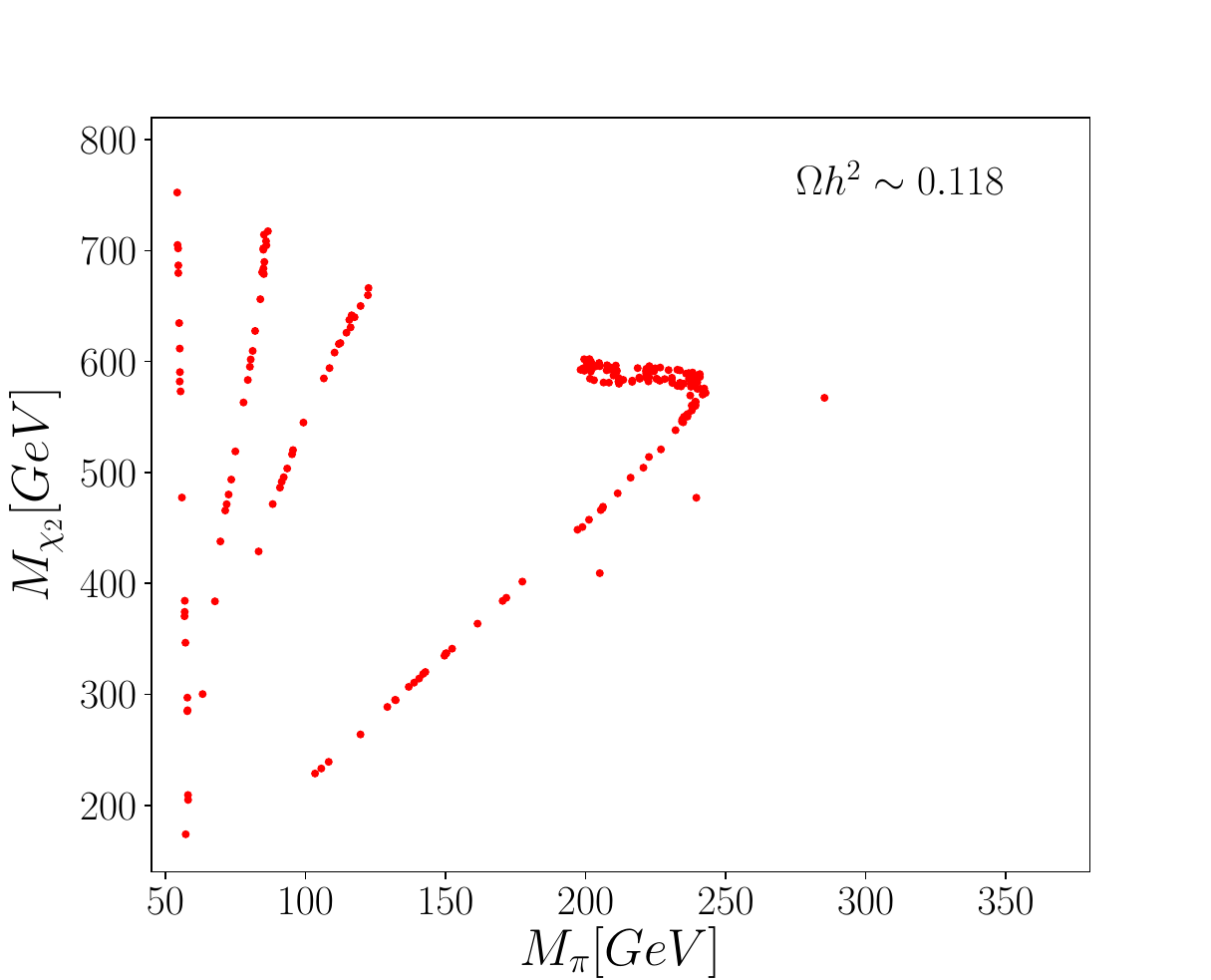}  
		\caption{}
		\label{fig:sub-4th1}
	\end{subfigure}
\caption{Set 1: In Fig (a) we use the $\Omega h^2 \leq 0.118$ for $\alpha_1=0.01$ radians, $\alpha_2=0.1$ radians and $\alpha_3=1.0$ radians showing the resonance effect (colour bar shows the relic density).  Fig (b) shows the allowed parameter space  for $\alpha_1=0.01$ radians, $\alpha_2=0.1$ radians and $\alpha_3=1.0$ radians  which satisfies $\Omega h^2 \sim 0.118$, direct and indirect detection constraints and collider constraints.}
\label{result_plot_case_1} 
\end{figure}
\begin{figure}{}
	\begin{subfigure}{.55\textwidth}
		\centering
		% include second image
		\includegraphics[width=1.1\linewidth]{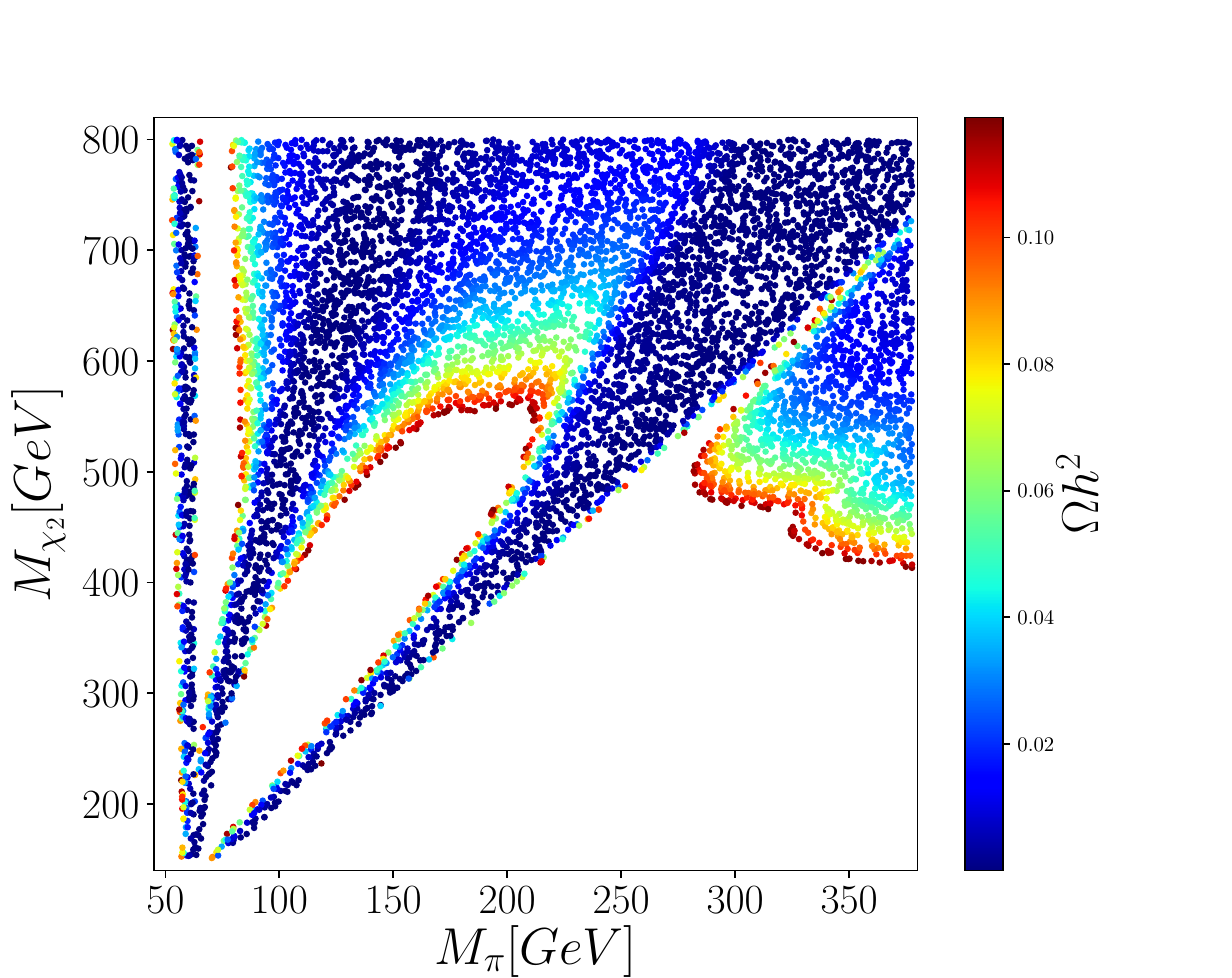}  
		\caption{}
		\label{fig:sub-5th}
	\end{subfigure}
	\begin{subfigure}{.5\textwidth}
		\centering
		% include second image
		\includegraphics[width=1.1\linewidth]{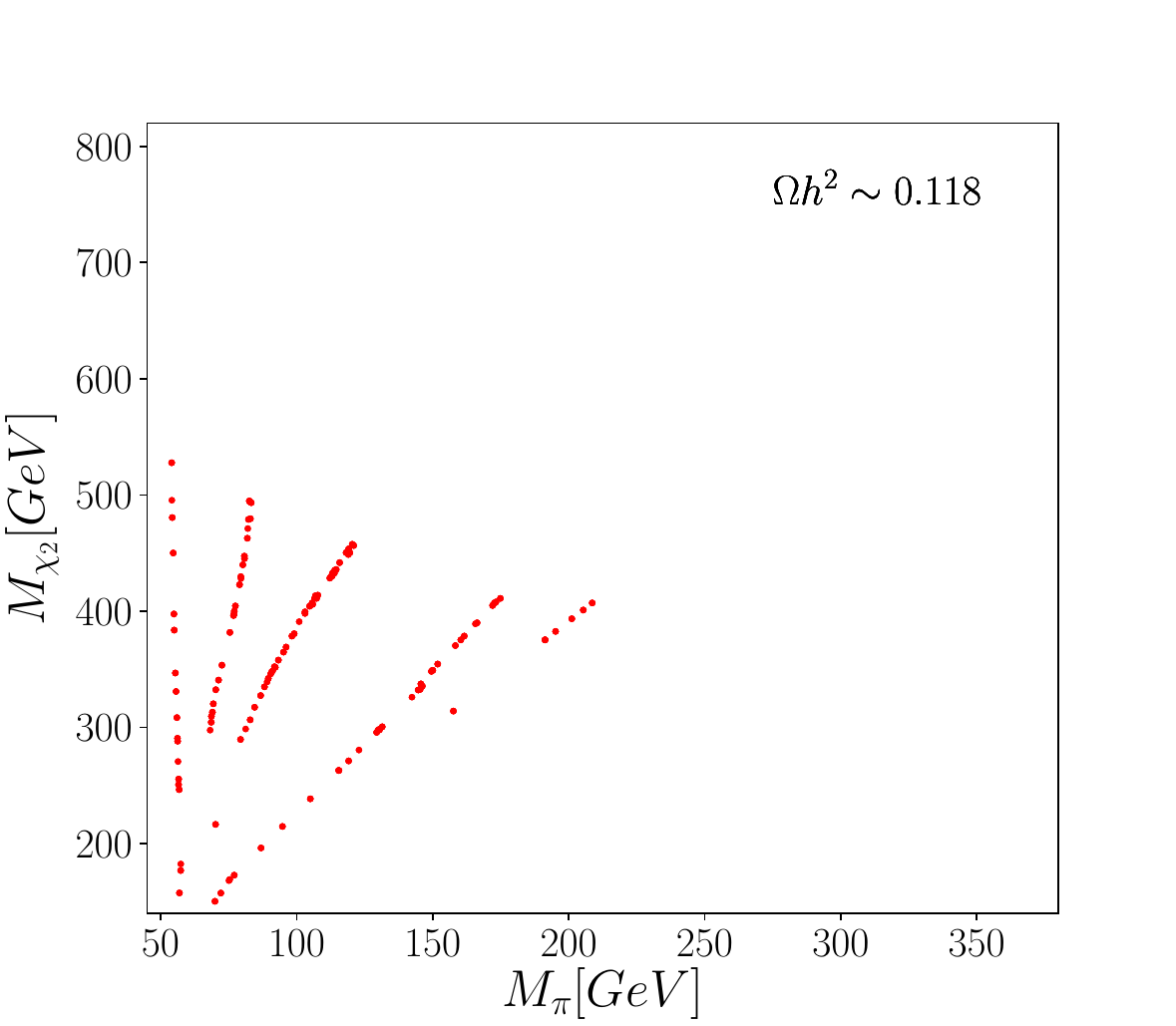}  
		\caption{}
		\label{fig:sub-5th1}
	\end{subfigure}
\caption{Set 2: Same as Fig. \ref{result_plot_case_1} with $\alpha_1=0.02$ radians, $\alpha_2=0.1$ radians and $\alpha_3=1.0$ radians.}
\label{result_plot_case_2}
\end{figure}
\begin{figure}{}	
		\begin{subfigure}{.53\textwidth}
		\centering
		% include second image
		\includegraphics[width=1.1\linewidth]{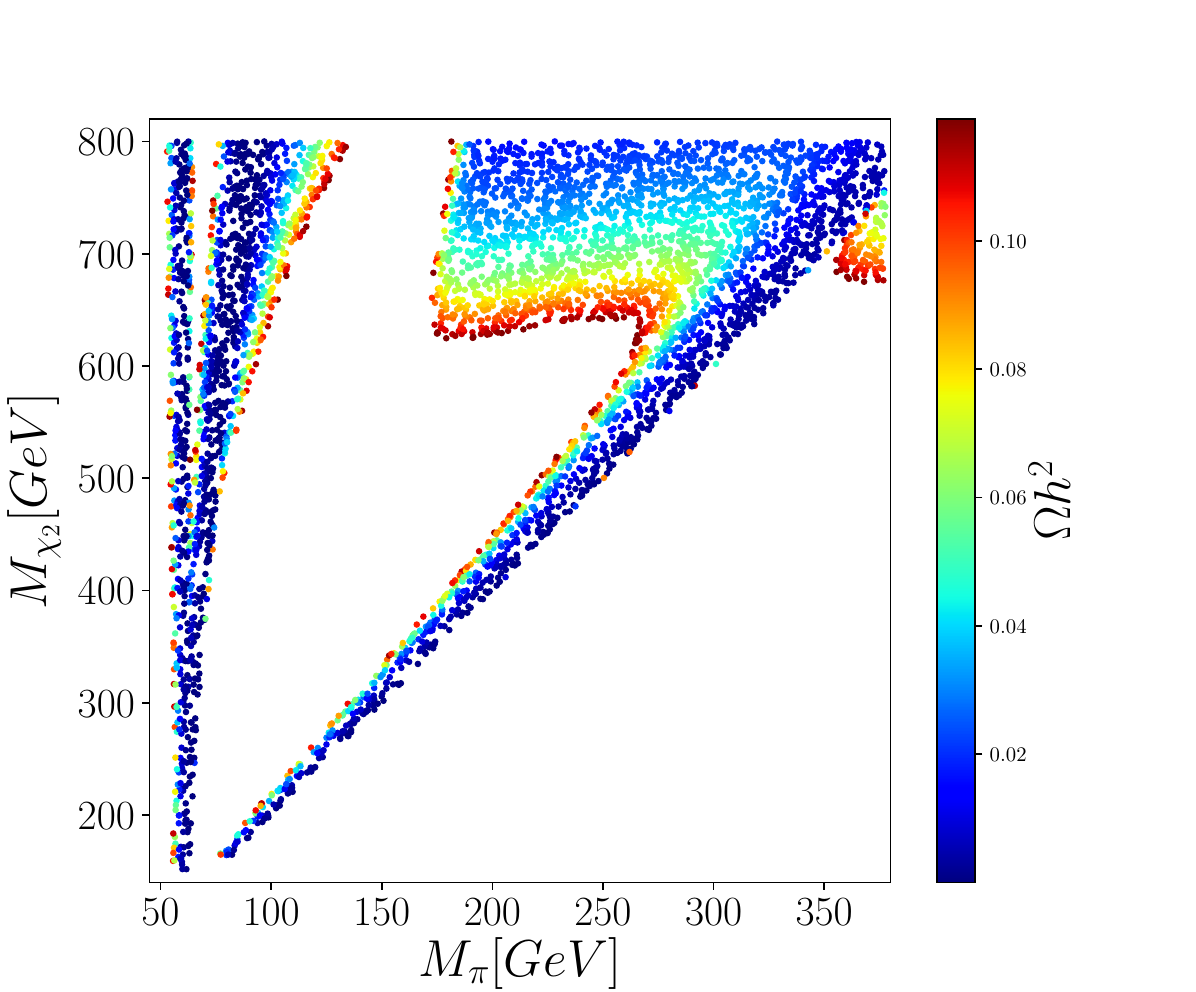}  
		\caption{}
		\label{fig:sub-6th}
	\end{subfigure}
	\begin{subfigure}{.5\textwidth}
		\centering
		% include second image
		\includegraphics[width=1.1\linewidth]{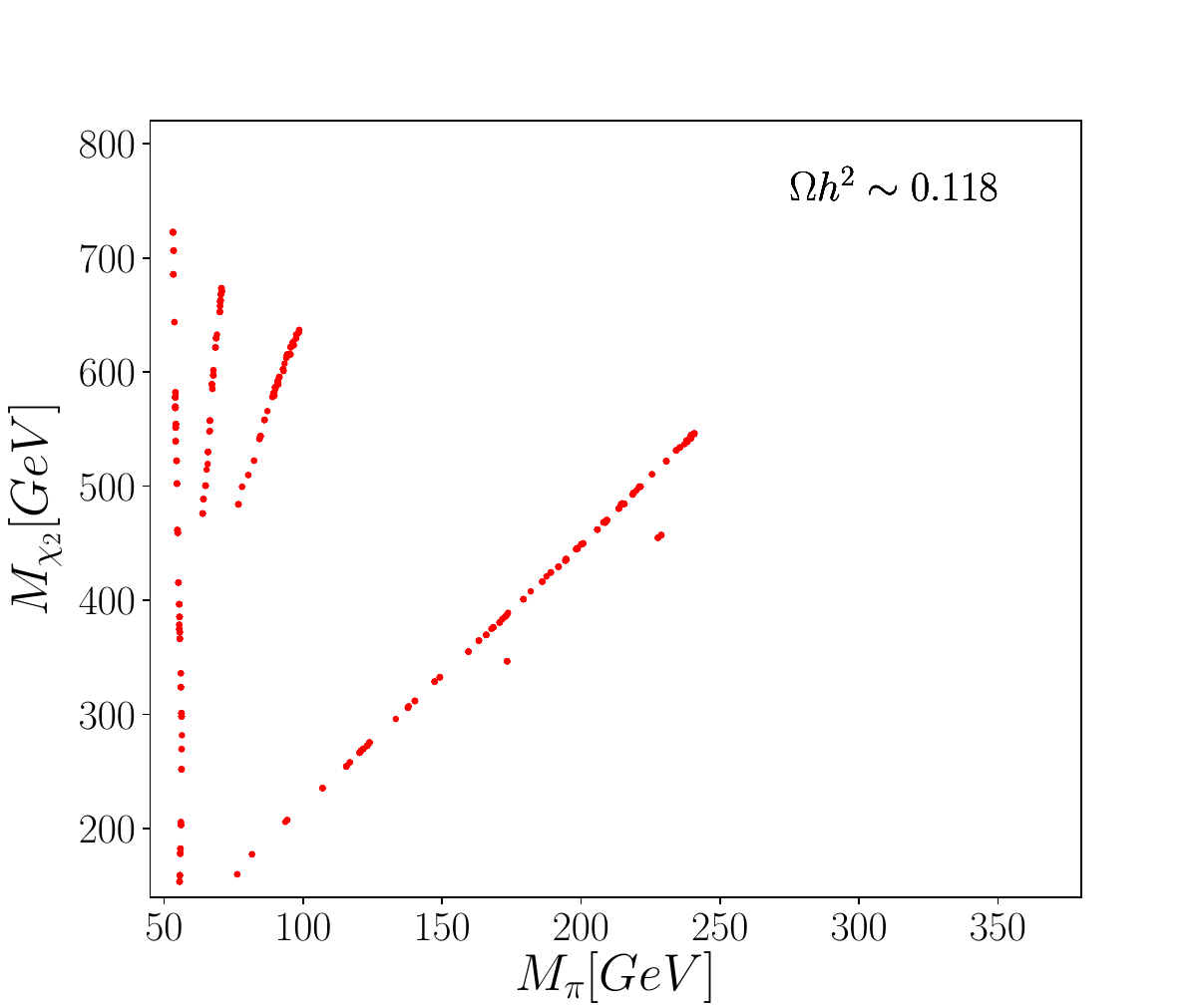}  
		\caption{}
		\label{fig:sub-6th1}
	\end{subfigure}
\caption{Set 3: Same as Fig. \ref{result_plot_case_1} with $\alpha_1=0.01$ radians, $\alpha_2=0.2$ radians and $\alpha_3=1.0$ radians.}
\label{result_plot_case_3}
\end{figure}
\begin{figure}{}
		\begin{subfigure}{.53\textwidth}
		\centering
		% include second image
		\includegraphics[width=1.1\linewidth]{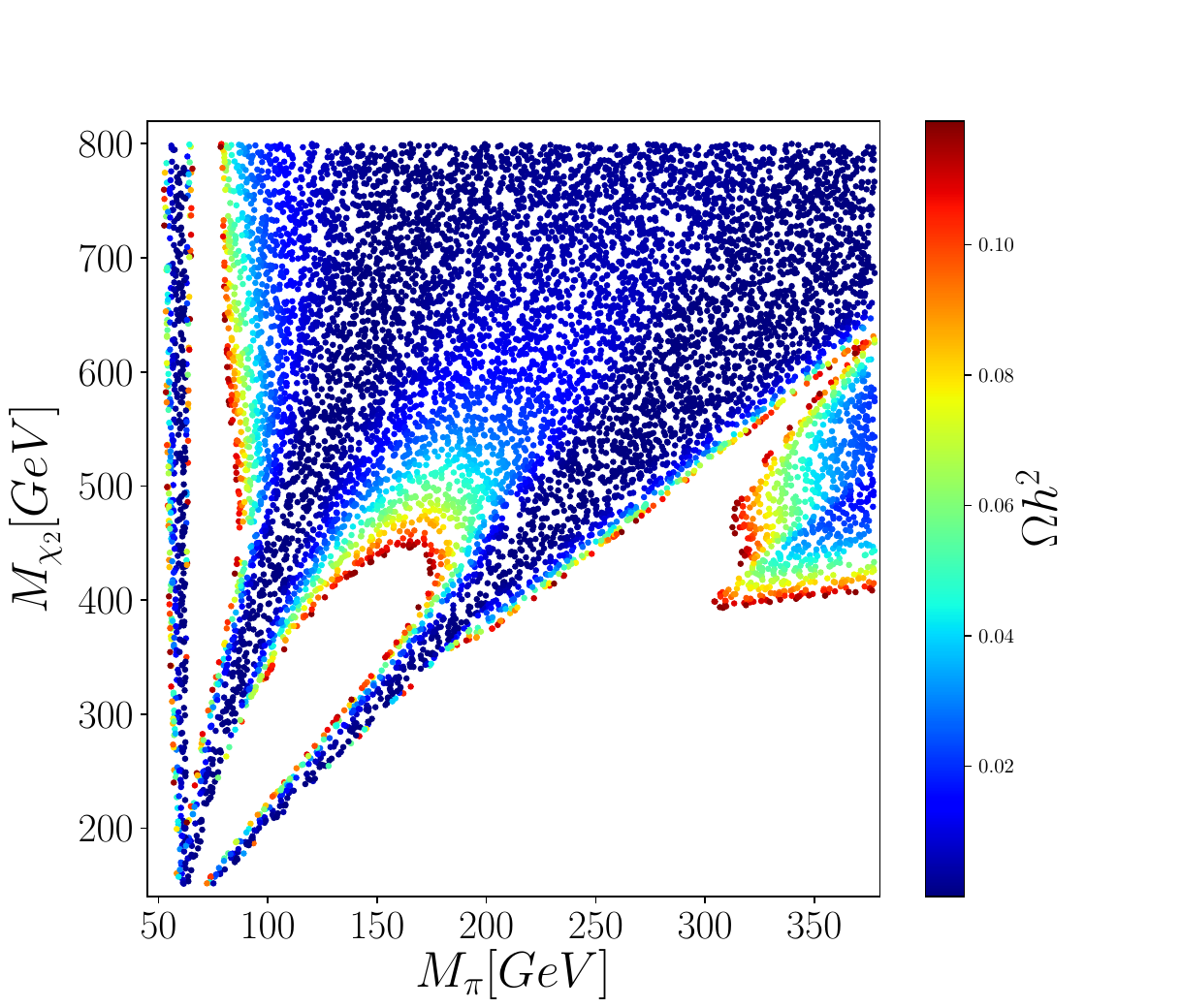}  
		\caption{}
		\label{fig:sub-7th}
	\end{subfigure}
	\begin{subfigure}{.5\textwidth}
		\centering
		% include second image
		\includegraphics[width=1.1\linewidth]{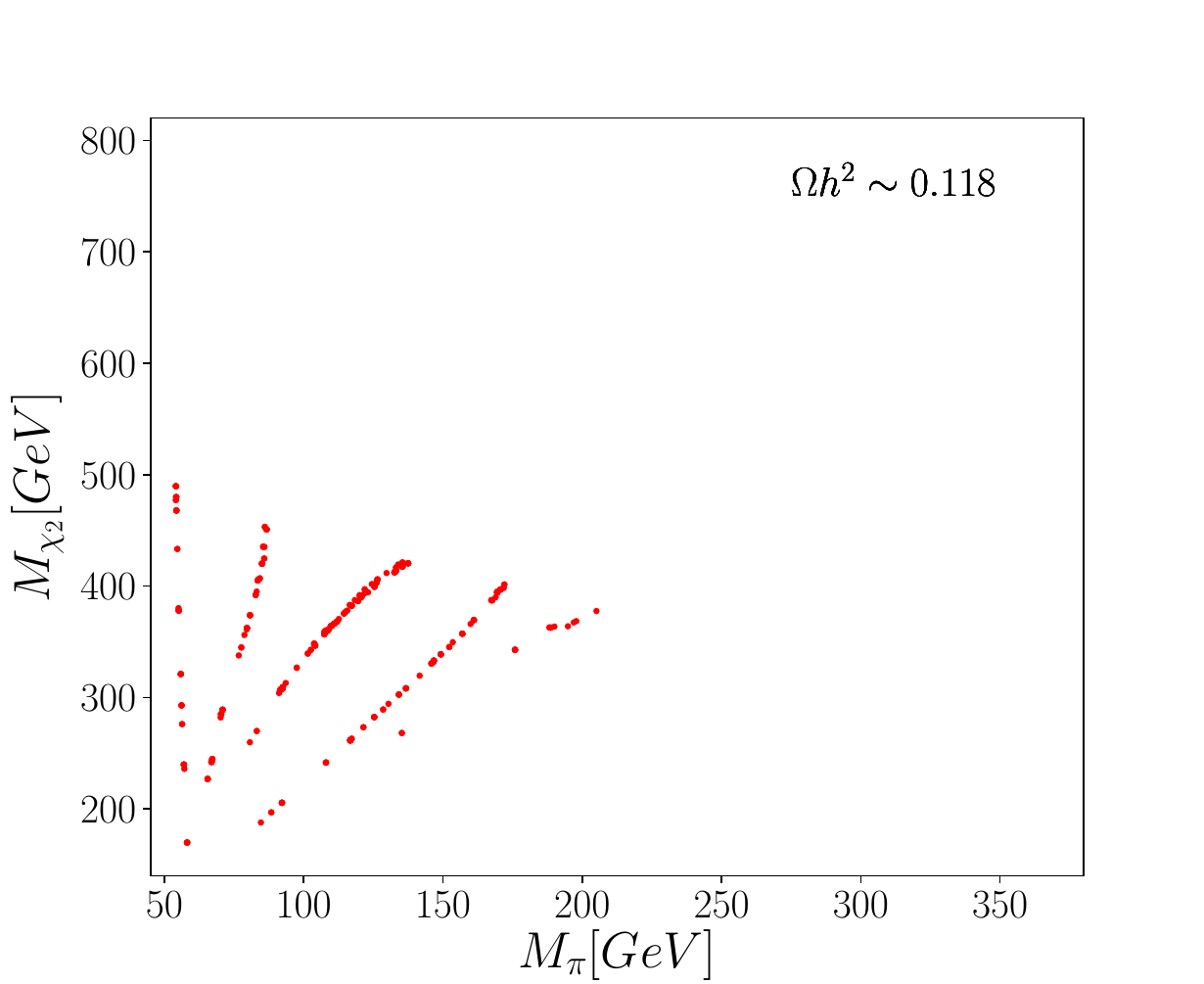}  
		\caption{}
		\label{fig:sub-7th1}
	\end{subfigure}
\caption{Set 4: Same as Fig. \ref{result_plot_case_1} with $\alpha_1=0.01$ radians, $\alpha_2=0.1$ radians and $\alpha_3=0.5$.
}   
\label{result_plot_case_4}
\end{figure}

\begin{figure}{}
\centering
  \includegraphics[width=0.6\linewidth]{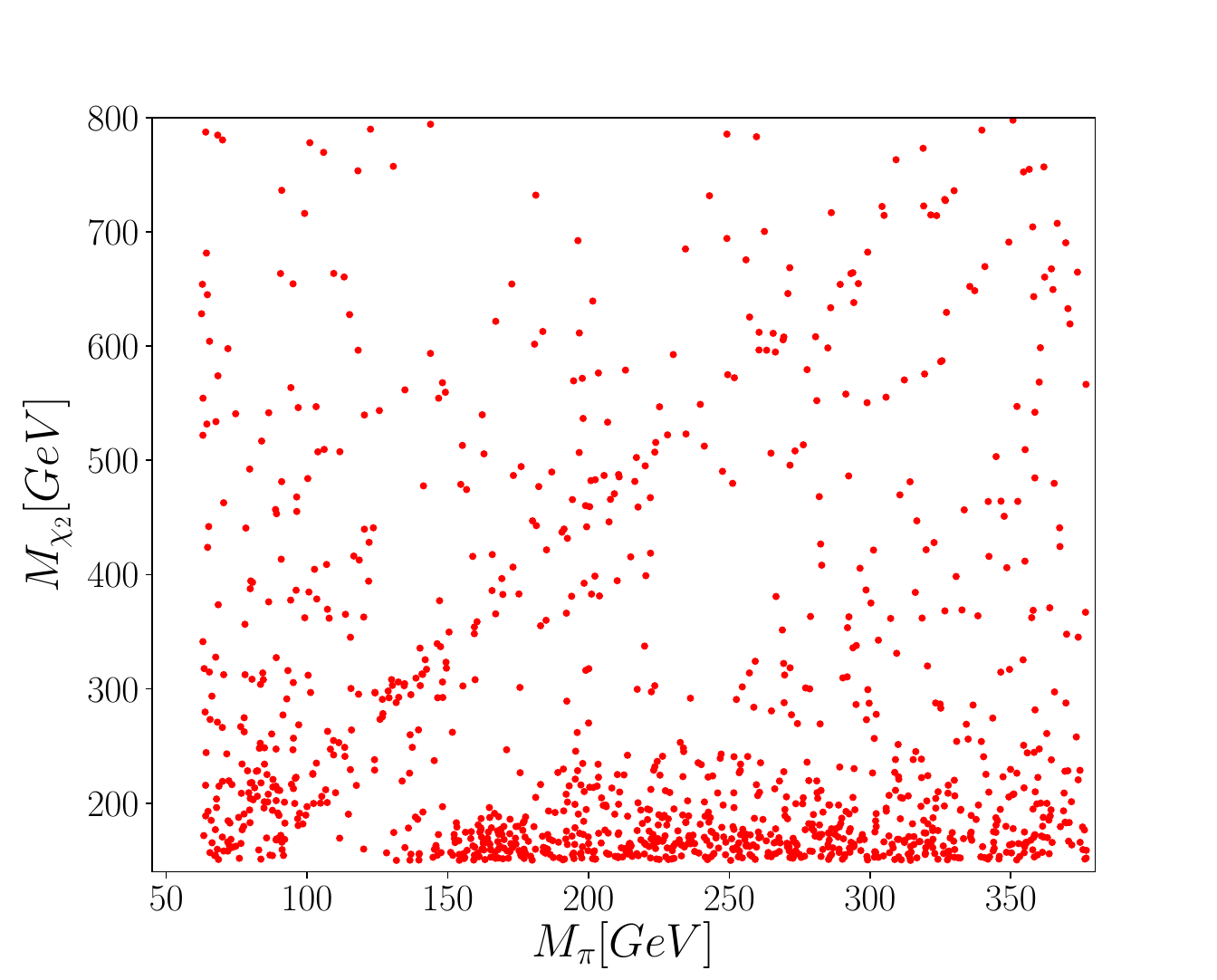}
  \caption{Allowed parameters in $M_\pi-M_{\chi_2}$ plane satisfying $\Omega h^2\sim0.118$, direct and indirect detection constraints and collider constraints. The mixing angles are chosen randomly over the range allowed by the collider constraint, as explained in text.}
  \label{finalplot}
\end{figure}

   The direct detection of dark matter with nucleons imposes a very strong constraint on the SI scattering cross-section of pions as seen in Fig. \ref{cross-section_plot}.  
This constraint is found to be much stronger in comparison with the indirect detection constraints. The relevant cross-sections for this case
are shown in Figs. \ref{bb}, \ref{tata},\ref{ww}, \ref{gam_gam} and \ref{gam_z}. These
 have peaks at the Breit-Wigner resonance of $\chi_1, \chi_2$ and $\chi_3$, but exactly at the resonance the pion relic density is significantly small and hence the indirect detection bounds at the resonance become
 unimportant \footnote{We should note that the direct detection and indirect detction constraint can be applied only when the relic density is $\sim 0.118$. For situations where $\Omega h^2_{\text{model}}<0.118$, one has to introduce the scale factor $\xi=\Omega h^2_{\text{model}}/\Omega h^2_{\text{CDM}}$ for direct detection and $\xi^2$ for indirect detection \cite{Casas:2017jjg}. In our case we used the direct and indirect detection constraint only for $\xi$ close to 1. }. 
These bounds are also  easily satisfied
 in the off-resonant region where the relic density is $\sim 0.118$ since
the cross-sections become very small. Hence we find that, once we apply 
the direct
detection constraints on the DM-nucleon cross-sections the indirect detection
constraints are
 automatically satisfied. We also see sudden dips in the cross section in Figs. \ref{bb}, \ref{tata}, \ref{ww}, \ref{gam_gam} and \ref{gam_z} for some values of the dark pion mass. These arise since
the coupling factors, such as $g_{\chi_2\pi\pi}$,  cross zero at these values of the dark pion mass. 
     
\begin{table}[htbp]
\begin{center}
\begin{tabular}{|c|c|c|c|}
\hline
\multirow{2}{3.2cm}{Benchmark Points}& \multicolumn{3}{p{3.5cm}|}{\centering Mixing Angles in radians} \\
\cline{2-4} & \multicolumn{1}{c|}{$\alpha_1$} & \multicolumn{1}{c|}{$\alpha_2$} & \multicolumn{1}{c|}{$\alpha_3$} \\ \hline
Set 1 & 0.01 & 0.1 & 1.0 \\
Set 2 & 0.02 & 0.1 & 1.0 \\
Set 3 & 0.01 & 0.2 & 1.0 \\
Set 4 & 0.01 & 0.1 & 0.5 \\
\hline
\end{tabular}
\end{center}
\caption{Different choices of mixing angles $\alpha_1$, $\alpha_2$ and $\alpha_3$.}
\label{Tab:SRNRValues}
\end{table}

%\begin{figure}
%  \centering
%  \includegraphics[width=0.75\linewidth]{relic_2.pdf}
%  \caption{$\alpha_1=0.02$ rad, $\alpha_2=0.2$ rad and $\alpha_3=1.2$ rad }
%  \label{fig:relic_2}
%\end{figure}

   In Figs. \ref{fig:sub-4th}, \ref{fig:sub-5th}, \ref{fig:sub-6th} and \ref{fig:sub-7th} we show the 
parameter space which satisfies $\Omega h^2 \leq 0.118$ 
for the choice of parameters corresponding to Set 1, 2, 3 and 4 respectively as given in Table \ref{Tab:SRNRValues}. 
The colour bar shows the relic density. 
We find that
there are three strips along the resonances arising due to exchange of
 particles 
 $\chi_1,\chi_2$ and $\chi_3$. In Figs. \ref{fig:sub-5th} and \ref{fig:sub-7th} we find that two of these
strips merge with one another. For these parameter sets (Set 2 and 4) we also find a much
larger number of points which deviate from resonance. The relic density 
falls sharply when the dark pion is at the resonance of any of the scalars,  
$\chi_1,\chi_2$ and $\chi_3$ (i.e. $M_\pi\sim M_{\chi_1,\chi_2,\chi_3}/2$). 
The red points are the parameters for which the relic density is close to 0.118.
 In Figs. \ref{fig:sub-4th1}, \ref{fig:sub-5th1}, \ref{fig:sub-6th1} and \ref{fig:sub-7th1} 
 we show the allowed parameter space in the $M_{\pi}$ and $M_{\chi_2}$ plane for parameter Set 1, 2, 3 and 4 (Table \ref{Tab:SRNRValues}) 
respectively  
which satisfies $\Omega h^2 \sim 0.118$ and direct detection constraint of XENON1T.  
We see that a considerable parameter range gets eliminated due to the direct detection constraints.
As mentioned earlier, constraints due to indirect detection are weaker and get
 automatically satisfied. 
In the final allowed parameter space (see Figs. \ref{fig:sub-4th1}, \ref{fig:sub-5th1}, \ref{fig:sub-6th1} and \ref{fig:sub-7th1}), 
we find that a large number of
points lie close to the resonance of the three scalar particles. However
in Fig. \ref{fig:sub-4th1}
we also see 
considerable parameter space for which $M_\pi$ deviates considerably 
from $M_{\chi_2}/2$. 

We next scan the entire parameter space corresponding to the three mixing angles and the masses $M_\pi$ and
$M_{\chi_2}$ by randomly selecting parameters over the range $\alpha_1, \alpha_2 \in [10^{-4},0.35]$ radians,
 $\alpha_3\in [10^{-2},1.55]$ radians, $M_{\chi_2}\in [150, 800]$ GeV
and $M_{\pi}\in [50, 400]$ GeV. The mixing angles are taken in this range since they automatically satisfy the
collider constraints, as shown in Fig. \ref{HS2}. After imposing all the remaining constraints, 
the final allowed 
values of $M_{\pi}$ and $M_{\chi_2}$ are shown in 
Fig. \ref{finalplot}.
We see that the model has sufficiently large allowed parameter space which 
satisfies all the phenomenological constraints. 
Some regions in the $M_{\pi}$ - $M_{\chi_2}$ plane are found to have higher density of points 
 in comparison to others. However, as we explain below, almost all of this parameter space is allowed.  
Some regions of this plot
arise due to contributions from close to resonant scattering
with s-channel exchange of $\chi_1$, $\chi_2$ and $\chi_3$  in similarity to 
 Figs. \ref{result_plot_case_1}, \ref{result_plot_case_2}, \ref{result_plot_case_3} and
 \ref{result_plot_case_4}.  
 As expected, 
 we do not get the right relic density exactly at resonance
but slightly away from it. 
For small $M_{\chi_2}$ we see a considerably high density of points.
These arise due to non-resonant contribution. We point out that  
the density of points is low near a resonance since in this case
the relic density shows a strong dependence on parameters 
and in order to satisfy all constraints 
the parameters need to be fine tuned.   
 This does not apply in non-resonant
regions where the cross section shows a relatively mild dependence 
on model parameters. The only forbidden regions are those corresponding to $M_\pi < 60$ GeV and a very narrow 
strip right along the resonance corresponding to any one of the three scalars. Other than that all regions 
are allowed. This applies even to some of the gaps seen in Fig. \ref{finalplot} for $M_\pi>60$ GeV. We have 
explicitly verified this by carefully exploring the parameter space corresponding to these regions.    

      As mentioned earlier our phenomenological analysis bears some resemblance to the Higgs portal DM, and a substantial analysis of Higgs portal DM is done before. The simplest yet most popular DM model of scalar singlet extension of SM is studied in depth for $\Omega h^2\leq0.118$ in \cite{Escudero:2016gzx, Casas:2017jjg} which rules out a vast region ($M_{\text{DM}}\leq500$ GeV) of parameter space because of strong direct and indirect detection constraints and Higgs invisible decay constraint. As mentioned in \cite{Casas:2017jjg}, results from XENON1T and LZ \cite{Akerib:2018lyp} might exclude a significantly large parameter space of the scalar singlet DM model apart from a very narrow region close to the Higgs resonance ($M_{\text{DM}}\sim M_{h_{\text{SM}}}/2$) \cite{Athron:2018ipf}. However   
our model naturally has a light dark pion and two more Higgs like scalars. 
Due to the presence of these heavier scalars $\chi_2$ and $\chi_3$ we find that 
a significant parameter space opens up both 
close to and away from the resonance of these
scalars.

We next consider the constraint given in Eq. \ref{eq:sigmaoverm}. 
For dark matter particles in the mass range considered in this section implies a very large DM-DM scattering cross section. For the
allowed parameter range we find that the largest value is many orders
of magnitude smaller than that given in Eq. \ref{eq:sigmaoverm}. This
is to be expected in any model based on the freeze out scenario and is only
possible in a freeze in scenario which allows for much smaller masses.

\subsection{Results: Freeze in Scenario}
\begin{figure}{}
		\centering
		% include first image
		\includegraphics[width=0.65\linewidth]{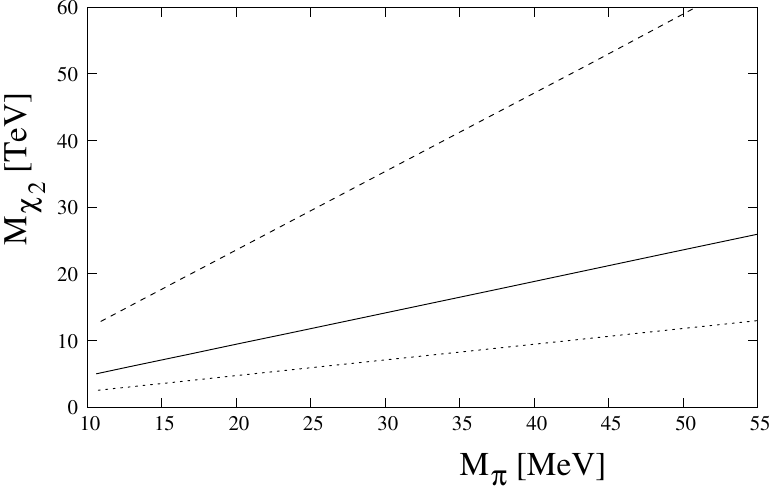}  
		\caption{The relationship between the dark pion mass $M_\pi$ and the $\chi_2$ meson mass $M_{\chi_2}$ for
the freeze in scenario. The short dashed, solid and long dashed lines correspond to parameter values  
($\lambda_6=0.5\times 10^{-4}$, $m/\eta = 0.5$),    
($\lambda_6=10^{-4}$, $m/\eta = 0.5$)   and  
($\lambda_6=10^{-4}$, $m/\eta = 0.2$) respectively.  
We have terminated the parameter range at
$M_\pi=55$ MeV. For larger values of this mass the strong sector coupling $\lambda_5$ becomes very large
and our calculation is not reliable.  }
		\label{fig:freezein}
\end{figure}

The basic formulas for the case of Freeze in scenario are given in section \ref{sec:freezein}. In Fig. \ref{fig:freezein} 
we show the relationship between the mass of dark pion and the dark sector particle $\chi_2$ for the allowed parameter
range which leads to the observed dark matter relic density and is also consistent
with the constraint on the dark matter scattering cross section
given in Eq. \ref{eq:sigmaoverm1}  for different values of the parameters $\lambda_6$ and $m/\eta$. Here we have set $\lambda=0.2$ 
and varied the parameter $\lambda_7$. Furthermore we have used Eqs. \ref{eq:relicFreezein}, \ref{eq:couplingFreezein} and \ref{eq:sigmaovermpi}
and have set the value of $\sigma_\pi/M_\pi$ equal to its upper limit given by Eq. \ref{eq:sigmaoverm1}.
Hence the value of $M_{\chi_2}$ can only take values below the lines shown in Fig. \ref{fig:freezein} for the corresponding
values of the parameters $\lambda_6$ and $m/\eta$. Larger values of this mass are ruled out by astrophysical observations. 
We point out that $M_{\chi_2}$ is directly proportional to the self coupling $\lambda_5$ and hence the region above the lines
shown in Fig. \ref{fig:freezein} correspond to values of $\lambda_5$ ruled out by observations.

In the limit $\lambda_5\lambda_6^2<<m^2/(2\eta^2)$ we find that the relationship between $M_\pi$ and $M_{\chi_2}$ is
approximately linear for fixed values of $\lambda_6$ and $m/\eta$, as seen in Fig. \ref{fig:freezein}.
 In this figure we have restricted
the dark pion mass $M_\pi$ to be less than 55 MeV. For larger values the strong sector coupling $\lambda_5$ becomes very
large and our leading order calculation of 
$\pi \,\pi$ scattering becomes unreliable. 
For a dark pion mass of 30 MeV we
find that $\lambda_5\approx 1.5$, which may be considered perturbative. 
We find that the experimental constraint on the Higgs decay to visible
sector particles is easily satisfied for all the values shown
in Fig. \ref{fig:freezein}.  
For this entire range of parameters 
the mass of dark sector particle $\chi_3$ is considerably larger than that of $\chi_2$.

\section{Conclusion}
\label{sec:conclusions}

We have studied 
the DM constraints within a conformal extension of the Standard Model
of particle physics. The model has strongly coupled dark sector which
triggers the electroweak symmetry breaking. 
The model leads to three neutral scalars $\chi_1$, $\chi_2$ 
and $\chi_3$ and we identify $\chi_1$ as the observed Higgs boson with mass 125 GeV. The model also predicts a pseudoscalar particle which we assume
is the DM candidate and refer to it as a dark pion. We perform all our
calculations using an effective Lagrangian for the strongly coupled dark 
sector. 
We first assume 
the freeze out scenario for obtaining the dark matter relic density. 
We find that the observed relic density $\Omega h^2=0.1187\pm 0.0017$
is obtained for a considerable range of parameter space,
some of which lies close to the resonance of any one of the neutral scalars. 
We find that the model is also able to explain
the constraints due to  
direct and indirect detection of DM and the collider constraints. 
The collider constraints impose limits on  the mixing angles of the neutral scalars. 
The upper bound on the SI DM-nucleon cross section obtained from direct detection experiments, such as, XENON1T,
 imposes a strong constraint on the model. 
The bounds arising due to searches of gamma ray 
signal from the galactic center and dSphs are found to be not very stringent  
and do not impose any additional constraint on the parameter space.
The dark pions have significant self interaction. However within the
freeze out scenario the value of $\sigma_{DM}/M_{\text{DM}}$ turns out to be relatively
small in comparison to the preferred astrophysical value. This is
 due to the large value of mass $M_{\text{DM}}$ required in this framework. 

We have also studied the implications of the conformal model assuming a freeze
in scenario for relic dark matter density. We argue that the model 
naturally leads to a very low mass dark pion while maintaining a relatively 
large mass scale of dark sector strong interactions. This is due to the
fact that the dark pion mass 
is controlled by chiral symmetry breaking terms in the dark sector. We can
choose this terms very small without fine tuning. 
 We find that
within the freeze in scenario the model is able to explain all astrophysical
and collider observations. As expected in this case we can also 
choose $\sigma_\pi/M_\pi$ close
to the value preferred by astrophysical observations.   

In our analysis we have assumed only
a single species of dark fermions which lead to only a single candidate
dark pion. Additional dark fermions and hence
dark pions can also be added in our model. This may 
open up additional regions 
in the parameter space.

\bibliographystyle{ieeetr}
\bibliography{2dm_ref}
\end{document}